\documentclass[11pt, oneside]{article}   	
\usepackage{geometry}                		
\usepackage{amsmath}				

\usepackage{algorithm}                              
\usepackage[noend]{algpseudocode}
\makeatletter
\def\BState{\State\hskip-\ALG@thistlm}
\makeatother
										
\usepackage{amssymb}
\usepackage{mathtools}
\usepackage{setspace}
\usepackage{fullpage}

\usepackage{rotating}
\usepackage{bm}                
\usepackage{dsfont}     
\usepackage{color}
\usepackage{graphicx}
\usepackage{url}
\usepackage{mathtools}
\DeclarePairedDelimiter{\ceil}{\lceil}{\rceil}              
\usepackage{titling}
\usepackage{titlesec}
\usepackage{authblk}
\usepackage{xr}
\usepackage{makecell}
\usepackage{bm}
\usepackage{cite}
\usepackage{hyperref}

\usepackage{caption}
\usepackage{subcaption}

\usepackage{float}

\newcommand{\hatW}{\hat{W}}
\newcommand{\hatbw}{\hat{\bw}}
\newcommand{\hatbG}{\hat{\bG}}
\newcommand{\hatbalpha}{\hat{\balpha}}
\newcommand{\be}{\bm{e}}
\newcommand{\expect}{\text{E}}
\newcommand{\hatG}{\hat{G}}

\newcommand{\bc}{\mathbf{c}}
\newcommand{\bx}{\mathbf{x}}
\newcommand{\bG}{\mathbf{G}}
\newcommand{\bw}{\mathbf{w}}
\newcommand{\balpha}{\bm{\alpha}}
\newcommand{\blambda}{\bm{\lambda}}

\newcommand{\beps}{\bm{\epsilon}}

\usepackage{etoolbox}
\makeatletter
\patchcmd{\@maketitle}{\raggedright}{\centering}{}{}
\patchcmd{\@maketitle}{\raggedright}{\centering}{}{}
\makeatother

\makeatother

\title{}
\title{A Bayesian method for estimating gene-level polygenicity under the framework of transcriptome-wide association study}
\author[1*]{Arunabha Majumdar}
\author[2*]{Bogdan Pasaniuc}
\affil[1]{Department of Mathematics, Indian Institute of Technology Hyderabad, Kandi, Telangana, India}
\affil[2]{Department of Pathology and Laboratory Medicine, University of California Los Angeles, California, USA}

\date{}

\begin{document}

\maketitle


*Correspondence: arun.majum@math.iith.ac.in (AM), pasaniuc@ucla.edu (BP)



\newpage

\section*{Abstract}

Polygnicity refers to the phenomenon that multiple genetic variants have a non-zero effect on a complex trait. It is defined as the proportion of genetic variants that have a nonzero effect on the trait. Evaluation of polygenicity can provide valuable insights into the genetic architecture of the trait. Several recent works have attempted to estimate polygenicity at the SNP level. However, evaluating polygenicity at the gene level can be biologically more meaningful. We propose the notion of gene-level polygenicity, defined as the proportion of genes having a non-zero effect on the trait under the framework of transcriptome-wide association study. We introduce a Bayesian approach {\em polygene} to estimate this quantity for a trait. The method is based on spike and slab prior and simultaneously provides an optimal subset of non-null genes. Our simulation study shows that {\em polygene} efficiently estimates gene-level polygenicity. The method produces downward bias for small choices of trait heritability due to a non-null gene, which diminishes rapidly with an increase in the GWAS sample size. While identifying the optimal subset of non-null genes, {\em polygene} offers a high level of specificity and an overall good level of sensitivity -- the sensitivity increases as the sample size of the reference panel expression and GWAS data increase. We applied the method to seven phenotypes in the UK Biobank, integrating expression data. We find height to be most polygenic and asthma to be the least polygenic. Our analysis suggests that both HDL and triglycerides are more polygenic than LDL. 

\clearpage

\section*{Introduction}

Beyond the discovery of genetic loci associated with a complex trait, it is crucial to estimate the overall distribution of the effect size of genome-wide genetic variants to understand the genetic architecture of the trait better \cite{zhang2018estimation}. Polygenicity is the proportion of genetic variants that have a nonzero effect on the trait. It is an essential characteristic of the effect size distribution and provides valuable insights into the genetic architecture of the trait. Efficient estimation of this quantity can help to improve the design of risk prediction models \cite{chatterjee2013projecting} or reveal the biological complexity of a trait \cite{o2019extreme}. Several approaches have attempted to estimate the polygenicity using its definition as the proportion of single nucleotide polymorphisms (SNPs), which have a nonzero effect on the trait under the framework of genome-wide association study (GWAS) \cite{zhang2018estimation, johnson2021estimation}. We refer to it as the SNP-level polygenicity. However, defining polygenicity at a gene level can be biologically more meaningful. In this paper, we propose the notion of gene-level polygenicity, which we define as the proportion of genes that have a nonzero effect on the trait. An expression data set containing data on the trait of interest would be ideal for estimating the gene-level polygenicity. However, most expression data sets have a limited sample size, and the data on various quantitative or disease traits may not be available.

Even though GWASs have discovered many common genetic variants associated with the risk of complex traits, most of these variants reside in non-coding genomic regions, making the functional interpretation of the GWAS signals challenging. Since the common variants associated with a complex trait tend to co-localize with expression quantitative trait loci (eQTLs) of the causal genes for the trait, a promising alternative approach is the transcriptome-wide association studies (TWAS) \cite{gamazon2015gene, gusev2016integrative}. TWAS integrates reference panel expression (and genotype) data with GWAS data to test for an association between the predicted genetic component of expression and a trait. Such studies combine eQTL effects estimated from the reference panel expression data and summary statistics from a GWAS. TWAS has identified numerous novel gene-trait associations and offers better biological interpretation than GWAS \cite{mancuso2018large}. Furthermore, it can be implemented efficiently using summary-level association data only \cite{gusev2016integrative}.

Considering the above deliberations, we develop a statistical approach to estimate the gene-level polygenicity under the framework of TWAS using summary-level association data. While evaluating SNP-level polygenicity, a crucial step is to account for the linkage disequilibrium (LD) among SNPs and identify the SNPs which originally have a nonzero effect \cite{zhang2018estimation, johnson2021estimation}. Two current methods, Genesis \cite{zhang2018estimation} and BEAVR \cite{johnson2021estimation}, estimate the number/proportion of susceptibility SNPs for a trait while taking into account LD structure. Genesis forecast the number of non-zero effect SNPs at a genome-wide level \cite{zhang2018estimation} which can be converted into a proportion of non-null SNPs. BEAVR partitions the genome into regions and estimates regional polygenicity \cite{johnson2021estimation}. A marginally associated SNP may not always have a nonzero effect. Instead, it may be in LD with an SNP, which has a nonzero effect.
Similarly, a marginally associated gene in TWAS may not originally have a nonzero effect. Instead, the marginal association may be due to a correlation between the predicted expression of the gene and another gene (gene co-regulation) where the latter truly has a nonzero effect. Such correlation between two genes' predicted expression can arise due to shared eQTLs or LD between eQTLs of the genes \cite{siewert2022leveraging}. We develop a Bayesian approach to estimate the proportion of genome-wide genes, the genetic component of expressions of which have a nonzero effect on the trait (non-null genes). We consider this proportion of non-null genes as a measure of gene-level polygenicity. We estimate the proportion of non-null genes explicitly accounting for correlation between the predicted genetic components of expressions. Our unified Bayesian framework also simultaneously provides an optimal subset of non-null genes. We refer to the method as {\em polygene} ({\em poly}genicity at {\em gene} level). We consider a continuous spike and slab prior to develop {\em polygene} \cite{george1993variable, ishwaran2005spike, rovckova2018bayesian, malsiner2018comparing}. The spike component represents the null effect, and the slab component represents the non-null effect. We perform a fully Bayesian inference based on MCMC while explicitly accounting for the covariance structure among the genes. Furthermore, the method uses marginal TWAS summary statistics, which are often publicly available. 

We perform extensive simulations to evaluate the performance of the approach. {\em polygene} efficiently estimates the gene-level polygenicity under various simulation scenarios. It produces a downward bias in the estimation when the heritability per non-null gene is small. The downward bias decreases rapidly as the GWAS sample size increases. Simulations also show that the q-value approach implemented in this context consistently produces a sizeable upward bias in all simulation scenarios. While identifying the optimal subset of non-null genes, {\em polygene} offers high specificity and good sensitivity across the simulation scenarios. The sensitivity improves with an increase in the sample size of the reference panel expression and GWAS data. We applied {\em polygene} to seven traits in the UK Biobank integrating expression data. Our analysis shows that height is the most polygenic, and asthma is the least polygenic. We also observe that both HDL and triglycerides are more polygenic than LDL.

\section*{Material and methods}

\subsection*{Overview of methods}

The standard TWAS consists of two-stage regressions. In the first stage, we consider a tissue of interest in the reference panel of expression (and genotype) data. We fit a penalized regression \cite{tibshirani1996regression, zou2005regularization} to evaluate the effect of genotypes of SNPs surrounding the gene (local SNPs) on the expression. From the regression for each gene, we obtain a prediction model to estimate the genetic component of the gene's expression based on its local SNPs. In the second stage, we use the prediction model to predict the genetic component of expression in the GWAS data based on the same set of local SNPs. We then regress a GWAS trait on the predicted expression to assess an association between the gene and the trait. We repeat this pipeline to obtain the marginal TWAS statistics for all the genes. Next, we derive the analytic formulas of the expectation vector and covariance matrix for the TWAS statistics of all genes. We assume a multivariate normal distribution of the TWAS statistics for the genes on each chromosome. Our main goal is to estimate the proportion of all such genome-wide genes which have a non-zero effect on the trait (gene-level polygenicity). We also aim to identify the optimal subset of non-null genes. To model sparsity, we consider a continuous spike and slab prior distribution \cite{george1993variable, malsiner2018comparing} for the TWAS effect sizes and develop a unified Bayesian approach to perform both the estimation of gene-level polygenicity and selection of non-null genes. With prior probability $p$, the TWAS effect size follows the slab distribution representing a non-null effect. In the data likelihood, we explicitly account for the covariance structure of genes. We derive the full-conditional posterior distributions of the model parameters to implement MCMC using the Gibbs sampling. Finally, we perform Bayesian inference based on the posterior sample of the model parameters obtained by the MCMC. 

\subsection*{Regression of gene expression on local SNPs in reference panel}

We describe the main steps of {\em polygene} for $m$ genes on a single chromosome, the extension of which for all chromosomes is straightforward. In the reference panel data, we regress the expression of $j^{th}$ gene on the genotypes of its local SNPs (e.g., SNPs within 0.5 MB window of the gene boundary), $j=1,\ldots,m$. We consider the following linear model:
\begin{equation}
 E_j = \bx'_j \bw_j + \epsilon_j
\end{equation}
$E_j$ denotes the mean-centered expression of $j^{th}$ gene in the specific tissue of interest. For the local SNPs, $\bx_j$ denotes the normalized genotype vector (centered for mean and then scaled by standard deviation) and $\bw_j$ denotes the effect size vector, $\epsilon_j$ denotes the random error. Since the number of local SNPs can be close to or larger than the sample size of the reference panel (e.g., GTEx data \cite{gtex2015genotype, gtex2017genetic}), we implement a penalized regression (e.g., Lasso \cite{tibshirani1996regression}, Elastic Net \cite{zou2005regularization}) to estimate $\bw_j$ individually for each $j=1,\ldots,m$.
 Suppose, we have $r_j$ local SNPs for $j^{th}$ gene. Let $r$ denote the total number of unique SNPs considered for the $m$ genes, and the SNPs are arranged in increasing order of base pair positions. Consider $r \times m$ matrix: $\hatW = [\hatbw_1, \ldots, \hatbw_m]$. For $j^{th}$ gene, all entries of $\hatbw_j$ are zero except for the $r_j$ local SNPs. Note that some of the $r_j$ entries can be zero due to fitting a penalized regression (e.g., Lasso). For each of these $m$ genes, we also assume that the local SNPs produced a significantly positive heritability of expression. We term such a gene as a locally heritable gene. We drop a gene from the downstream analysis if it is not locally heritable.

\subsection*{Regression of trait on predicted expression in GWAS data}

In general, expression measurements are not available in GWAS data. Suppose the genotype data for the set of local SNPs considered for a gene in the reference panel is also available in GWAS data. In that case, we can predict the genetic component of the gene's expression using $\hatW$ (obtained in the first stage regression). We estimate the genetic component of expression for $m$ genes as: $\hatbG = X \hatW$. $X_{n \times r}$ is the genotype matrix for $n$ individuals and $r$ local SNPs in the GWAS data. The genotype data of each SNP is normalized to have zero mean and unity variance. Subsequently, we can perform a multiple linear regression of a continuous GWAS trait $Y$ on $\hatbG$ to evaluate the joint effect of the $m$ genes on $Y$ as:
\begin{equation}
\label{eqn:joint-regress-Y-Ghat}
 Y = \hatbG \balpha + \be
\end{equation}
$Y$ denotes the trait vector for $n$ individuals. $\balpha = (\alpha_1,\ldots,\alpha_m)^{'}$ denotes the joint effect sizes for the $m$ genes. We ignore the intercept term considering $Y$ to be mean-centered. We note that $\hatbG$ is an estimate of true $\bG$ and involves some uncertainty (due to variability in $\hatW$) which is ignored in a standard TWAS. Here, $\be$ denotes the error term and we assume that $\expect(\be) = \bm{0}, \text{cov}(\be) = \sigma^2_e I_{n \times n}$. The joint ordinary least square (OLS) estimate of $\balpha$ is given by: $\hatbalpha = (\hatbG'\hatbG)^{-1}\hatbG'Y$. We note that an OLS estimate of $\balpha$ should be reliable, because the number of locally heritable genes on a single chromosome (e.g., 500) is expected to be much smaller than the sample size of a contemporary GWAS data (e.g., 10,000). In the standard TWAS, we consider a univariate regression: $E(Y) = \hatG_j \gamma_j$, where $\hatG_j$ denotes the predicted genetic component of $j^{th}$ gene's expression ($j^{th}$ column of $\hatbG$). We test for a marginal association, $H_0: \gamma_j = 0$ vs $H_1: \gamma_j \neq 0$. OLS estimate of $\gamma_j$ is given by: $\hat{\gamma}_j = (\hatG_j^{'} \hatG_j)^{-1} \hatG_j^{'} Y = \frac{\hatG_j^{'} Y}{\hatG_j^{'} \hatG_j}$ with s.e.$(\hat{\gamma}_j) = \frac{ {\hat{\sigma}}^2_y }{\sqrt{\hatG_j^{'} \hatG_j}}$. Assuming that $Y$ is normalized to have a mean zero and variance one, the $Z$ statistic for testing the marginal association is given by: $z_j = \frac{\hatG_j^{'} Y}{\sqrt{\hatG_j^{'} \hatG_j}}$. We aim to develop our method based on marginal TWAS statistics which are often publicly available.

\subsection*{Expectation of the vector of marginal TWAS statistics}

We derive the expectation of marginal TWAS statistics considering the joint model of $Y$ in equation \ref{eqn:joint-regress-Y-Ghat}: $E(Y)=\hatbG \balpha$. Thus,
\begin{equation}
\expect(z_j) = \frac{\hatG_j^{'} \expect(Y)}{\sqrt{\hatG_j^{'} \hatG_j}} = \frac{\hatG_j^{'} \hatbG \balpha}{\sqrt{\hatG_j^{'} \hatG_j}}
\end{equation}
Define the diagonal matrix $M_1=diag(\frac{1}{\sqrt{\hatG_1^{'} \hatG_1}}, \ldots, \frac{1}{\sqrt{\hatG_m^{'} \hatG_m}})$. $\hatbG' \hatbG = \hatW^{'}X^{'}X\hatW = n \hatW^{'}V\hatW$. $V=\frac{1}{n}X^{'}X$ is the LD matrix for the $r$ SNPs in GWAS data. Considering the vector of marginal $Z$ statistics across $m$ genes simultaneously, we obtain that:
\begin{equation}
   \expect(Z) = M_1 \hatbG' \hatbG \balpha = n M_1 \hatW^{'}V\hatW \balpha 
\end{equation}
Since $\hatbG' \hatbG = n \hatW^{'}V\hatW$, $j^{th}$ diagonal element of $\hatbG' \hatbG$ is given by: $\hatG_j^{'} \hatG_j = n [\hatW^{'}V\hatW]_{jj}$. We define a diagonal matrix, the diagonal entries of which are the diagonal elements of $\hatW^{'}V\hatW$: $diag(\hatW^{'}V\hatW)$ = $diag([\hatW^{'}V\hatW]_{11}, \ldots, [\hatW^{'}V\hatW]_{mm})$. Thus, $M_1 = \frac{1}{\sqrt{n}} \sqrt{\{diag(\hatW^{'}V\hatW)\}^{-1}}$. Hence,
 
\begin{equation}
\expect(Z) =   \sqrt{\{diag(\hatW^{'}V\hatW)\}^{-1}} \mbox{ } \hatW^{'}V\hatW (\sqrt{n}\balpha)
\end{equation}
We denote $\blambda = \sqrt{n} \balpha$, $M_2 = \sqrt{\{diag(\hatW^{'}V\hatW)\}^{-1}}$, and $S =  \hatW^{'}V\hatW$. Thus,
\begin{equation}
\expect(Z) = M_2 S  \blambda
\end{equation} 
We note that $\alpha_j=0$ is equivalent to $\lambda_j=0$, $j=1,\ldots,m$.

\subsection*{Covariance matrix of the vector of marginal TWAS statistics}

We have $Z = \frac{1}{\sqrt{n}} M_2 \hatbG'Y = \frac{1}{\sqrt{n}} M_2 \hatbG' (\hatbG \balpha + \hatbG \beps)$. It can be derived that:
\begin{equation}
  \text{cov}(Z) = M_2 S M_2, \text{ where } M_2 = \sqrt{\{diag(\hatW^{'}V\hatW)\}^{-1}} \text{ and } S = \hatW^{'}V\hatW
\end{equation}
We note that for a vector of $Z$ statistics, the correlation matrix is the same as the covariance matrix. The full covariance matrix for all chromosomes will be a block-diagonal matrix, where each block corresponds to each chromosome. The inverse of a block-diagonal matrix will again be a block-diagonal matrix containing the corresponding inverted sub-matrices. We also note that two genes on a single chromosome residing far apart (e.g., $> 2$ MB) are likely to have a zero correlation. This induces a sliding-window type of covariance structure for the genes on each chromosome. 

\subsection*{Continuous spike and slab prior}

Our main goal is simultaneously estimating the gene-level polygenicity and the optimal subset of non-null genes under a unified Bayesian framework. We consider a Bayesian approach based on continuous spike and slab prior (\cite{george1993variable, malsiner2018comparing}), where the spike component represents a null effect, and the slab component represents a non-null effect. As mentioned above, we first describe the method for a set of $m$ genes on a single chromosome which can easily be extended to all chromosomes. From previous sections we have: $\text{E}(Z) = M_2 S \blambda$ and $\text{cov}(Z) = M_2 S M_2$, where $M_2 = \sqrt{\{diag(\hatW^{'}V\hatW)\}^{-1}}$ and $S =  \hatW^{'}V\hatW$. Note that the $Z$ statistic for each gene follows normal. We assume that the vector of $Z$ statistics follows a multivariate normal distribution.
\begin{equation}
Z|\blambda \sim N_m(M_2 S \blambda, M_2 S M_2)
\end{equation}
We consider a continuous spike and slab prior distribution for $\blambda$, the vector of TWAS effect sizes for $m$ genes. For $j=1,\ldots,m$, we consider the prior of $\lambda_j$ as:
\begin{equation} \label{eq:spike-slab}
\begin{split}
\lambda_j | c_j, \nu & \sim (1-c_j) N(0, \sigma_0^2) + c_j N(0, \nu^2 \sigma_0^2); \mbox{ } \nu >> 1,  \mbox{ } \sigma_0^2 = 10^{-10} \\
 & \text{P}(c_j=1|p) = p, \mbox{   } \text{P}(c_j=0|p) = 1-p \\
 & \mbox{   } p | a_1,a_2 \sim \text{Beta}(a_1,a_2) \\
 &\nu^2 \sigma_0^2 = \sigma_1^2 \sim \text{Inverse Gamma}(b_1,b_2)
\end{split}
\end{equation}
The latent variable $c_j$ defines whether the $j^{th}$ gene has a non-zero effect or not. When $c_j=0$, $\lambda_j \sim N(0, \sigma_0^2)$, and when $c_j=1$, $\lambda_j \sim N(0,\nu^2 \sigma_0^2)$. We consider a very small fixed value of $\sigma_0^2 = 10^{-10}$ and a relatively much larger value of $\sigma_1^2 = \nu^2 \sigma_0^2$ (e.g., $10$) such that $\nu = \frac{\sigma_1}{\sigma_0} >> 1$. Since $\sigma_0^2$ is nearly zero, if $c_j=0$, $\lambda_j$ would be very small which can safely be considered as zero ($j^{th}$ gene is null). We can consider a sufficiently large value of $\sigma_1^2$ such that, if $c_j=1$, $\lambda_j$ can be treated as non-zero ($j^{th}$ gene is non-null). We refer to $\sigma_0^2$ as the spike variance and $\sigma_1^2$ as the slab variance. A random variable following N$(0,10^{-10})$ has $99\%$ probability of taking a value between $(-2.6 \times 10^{-5}, 2.6 \times 10^{-5})$. The proportion of non-null genes, $p$, measures the gene-level polygenicity. The vector of latent variables, $C = (c_1,\ldots,c_m)$, defines the subset of non-null genes which have a non-zero effect on the trait.

For convenience we assume that $\lambda_1,\ldots,\lambda_m$ are independently distributed in the prior. Conditioned on $p$, the configuration indicators $(c_1,\ldots,c_m)$ are i.i.d. Bernoulli$(p)$, where $p \sim$ Beta$(a_1,a_2)$. We assume that $\sigma_1^2$ $\sim$ Inverse Gamma$(b_1,b_2)$. We discuss a method of moments approach to choose the hyperparameters $(a_1,a_2)$ and $(b_1,b_2)$ in a later section. We perform a fully Bayesian inference on the gene-level polygenicity and the optimal subset of non-null genes. Note that when $\sigma_0^2 = 0$, the prior has a positive mass at $\lambda_j=0$, and is known as Dirac spike and slab prior (\cite{malsiner2018comparing}). 
  
\subsection*{MCMC}

We implement the Markov Chain Monte Carlo (MCMC) algorithm using Gibbs sampling to generate a posterior sample of the model parameters. We provide the full-conditional posterior distributions in the following. In a given MCMC iteration, we update $\blambda, C, p, \sigma_1^2$ sequentially.

\vspace{0.4cm}

\noindent {\it Full-conditional posterior distribution of $\blambda$}:
Define $D_C$ to be a diagonal matrix with its $j^{th}$ diagonal element defined in the following way: if $c_j=0$, it is chosen as $\sigma_0^2$ and if $c_j=1$, it is chosen as $\sigma_1^2$. Thus, the diagonal entries of $D_C$ are $(\sigma_{c_1}^2, \ldots, \sigma_{c_m}^2)$. We obtain the full-conditional posterior distribution of $\blambda$ as:
\begin{equation*}
\blambda|Z, C, p, \sigma_1^2 \sim N_m(\bm{\mu}, \Sigma), \text{ where } \Sigma = (S+D_{\bc}^{-1})^{-1} \text{ and } \bm{\mu} = \Sigma M_2^{-1} Z  
\end{equation*}
\noindent {\it Full-conditional posterior distribution of $C$}:
Let $C_{-j} = (c_1,\ldots,c_{j-1},c_{j+1},\ldots,c_m)$. We update the configuration indicators using the full-conditional distribution of $c_j, j=1,\ldots,m$.
\begin{equation*}
 P(c_j=0|C_{-j}, Z, \blambda, p, \sigma_1^2) = \frac{1}{1+\frac{p}{1-p} \frac{f(\lambda_j|c_j=1)}{f(\lambda_j|c_j=0)}}
\end{equation*}
 Here, $f(\lambda_j|c_j=1) = N(\lambda_j;0,\sigma_1^2)$ and $f(\lambda_j|c_j=0) = N(\lambda_j;0,\sigma_0^2)$. $N(x;\mu,\sigma^2)$ denotes the normal density at $x$ given  $\mu,\sigma^2$ to be the mean and variance. $P(c_j=1|C_{-j}, Z, \blambda, p, \sigma_1^2) = 1-P(c_j=0|C_{-j}, Z, \blambda, p, \sigma_1^2)$.

\vspace{0.4cm}
 
\noindent {\it Full-conditional posterior distribution of $p$}:
Denote $m_1= \# \{c_j | c_j=1, j=1,\ldots,m\} = \sum_{j=1}^{m} c_j$ and $m_0=m-m_1$. We update $p$ from: $p|Z, \blambda, C, \sigma_1^2 \sim \text{Beta}(a_1+m_1,a_2+m_0)$. 

\vspace{0.4cm}

\noindent {\it Full-conditional posterior distribution of $\sigma_1^2$}:
Next, we update the slab variance $\sigma_1^2$ using the following Inverse-Gamma distribution: $\sigma_1^2 | Z, \blambda, C, p \sim \text{Inverse-Gamma}(b_1+\frac{m_1}{2},b_2+\frac{1}{2} \sum_{{\substack j:c_j=1}} \lambda_j^2)$. Note that, if $m_1=0$, $\sigma_1^2$ is updated from its prior, Inverse-Gamma$(b_1,b_2)$. 

\subsection*{Posterior inference}
 
After a certain burn-in period of the MCMC, we collect the posterior sample of model parameters. We obtain various posterior summaries based on the MCMC sample. We use the posterior median of $p$ as the point estimate of the gene-level polygenicity. We consider a $5\%-95\%$ central posterior interval of $p$ to evaluate the uncertainty in the estimate. In a given MCMC iteration, the vector of latent variables $C=(c_1,\ldots,c_m)$ defines the subset of non-null genes. For $j=1,\ldots,m$, $c_j=1$ implies that the $j^{th}$ gene is included in the subset, and $c_j=0$ implies that it is excluded. We consider the subset of non-null genes, observed with the maximum frequency in the posterior sample, as an estimate of the optimal subset of non-null genes (maximum a posteriori (MAP) estimate).

\subsection*{Choice of the hyperparameters}

We adopt a method of moments (MOM) approach to choosing the hyperparameters in the prior distributions of $p$ and $\sigma_1^2$. To estimate gene-level polygenicity, we focus on the TWAS data obtained from the tissue type that is most relevant for the trait. However, if expression data are available for other tissues in the reference panel (e.g., GTEx data), we can obtain TWAS statistics for different tissue types. We consider the trait's TWAS data from a closely relevant tissue to estimate the hyperparameters. For example, if we consider the brain the primary tissue type of interest for BMI, we can use the adipose-specific TWAS statistics to estimate the hyperparameters. This is a partially empirical Bayes approach because we use the same GWAS data while computing the TWAS statistics for both the tissue types.

The two shape parameters in the Beta prior of $p$ are $a_1,a_2$, and the shape and scale parameters in the inverse-gamma prior of $\sigma_1^2$ are $b_1,b_2$. For $j^{th}$ gene, we integrate out $\lambda_j$ to obtain the distribution of the TWAS statistics conditioned only on $p,\sigma_1^2$: $z_j|p,\sigma_1^2 \sim p N(0,1+\sigma_1^2) + (1-p) N(0,1)$. Note that the variance of $z_j$ corresponding to the spike component is $(1+10^{-10})$ which we approximate as $1$. We can derive the $k^{th}$ order raw population moment as $E(z_j^k) = E_p E(z_j^k | p)$, where $E(z_j^k | p) = E_{\sigma_1^2}E(z_j^k | p, \sigma_1^2)$. Here, $E_{\sigma_1^2}E(z_j^k | p, \sigma_1^2) = p E(z_j^k|z_j \sim N(0,1+\sigma_1^2)) + (1-p) E(z_j^k|z_j \sim N(0,1))$. The odd order moments of $z_j$ are zero, because the distribution of $z_j$ is symmetric at zero conditioned on $p,\sigma_1^2$. Thus, $E(z_j) = E(z_j^3) = 0$. We obtain the second and fourth order moments of $z_j$ as the following:
\begin{equation*}
E(z_j^2) = 1 + \frac{a_1}{a_1 + a_2} \frac{b_2}{b_1-1}
\end{equation*}    
\begin{equation*}
E(z_j^4) = 3 [ \frac{a_2}{a_1 + a_2} + \frac{a_1}{a_1 + a_2} (1 + \frac{2b_2}{b_1-1} + \frac{b_2^2}{(b_1-1)(b_1-2)}) ]
\end{equation*} 
Using the MOM approach we equate $E(z_j^2) = m_2^{'}$ and $E(z_j^4) = m_4^{'}$, where $m_2^{'}$ and $m_4^{'}$ are the second and fourth order sample raw moments: $m_2^{'} = \frac{1}{M}\sum_{j=1}^{M} z_j^2$ and $m_4^{'} = \frac{1}{M}\sum_{j=1}^{M} z_j^4$, where $M$ is the total number of genes on all chromosomes. Since we have two equations in four unknowns, we fix the values of $a_1$ and $b_1$ as: $a_1=0.1$, $b_1=3$. Then we solve the equations to obtain the choices of $a_2$ and $b_2$. Note that it is challenging and tedious to obtain the sixth and eighth order moments and finally solve for all the four unknowns.

\section*{Results}

\subsection*{Simulation study}

We perform an extensive simulation study to evaluate the efficiency of {\em polygene} concerning estimating the gene-level polygenicity and the optimal subset of non-null genes. We use the actual genotype data of $337K$ white-British individuals in the UK Biobank (UKBB) for our simulations.

\subsubsection*{Simulation design}

The software package Fusion \cite{gusev2016integrative} analyzed the expression-genotype data in the Young Finish Sequencing (YFS) study to identify all the locally heritable genes in the whole blood tissue. Out of 4700 locally heritable genes, we considered a subset of 2988 genes located in chromosomes 7-22 for our simulations. For each gene, we consider the same set of local SNPs that Fusion included to estimate the local heritability and the prediction model of genetic component of expression.
We consider a subset of $n_E$ (e.g., 1000) individuals randomly drawn from the UKBB individuals as the reference panel of expression data. For each gene, we use a linear model to simulate the expression in the reference panel: $E_j = \bx'_j \bw_j + \epsilon_j$. Here $E_j$ denotes the expression of the $j^{th}$ gene, $\bx_j$ denotes the genotypes of the set of local SNPs, and $\bw_j$ denotes the corresponding effect sizes. Under the assumption that $\text{V}(E_j)=1$, we consider $\epsilon_j \sim N(0,1 - h^2_{ej})$, where $h^2_{ej}$ is the local heritability of the expression of $j^{th}$ gene (due to the local SNPs). We assume that a proportion of the local SNPs affect the expression. If $r_{cj}$ denotes the number of such SNPs, each element of $\bw_j$ follows $N(0,\frac{h^2_{ej}}{r_{cj}})$. While simulating the expression, we standardize the genotype data of each local SNP to have zero mean and unity variance. We apply Fusion \cite{gusev2016integrative} to simulated data to identify the genes with significant local heritability using a p-value threshold $\frac{0.05}{20000}$. To obtain the prediction model of genetic component of expression, we implement the penalized regression using the elastic net penalty \cite{zou2005regularization} for the expression and local SNPs' genotypes. We only consider the locally heritable genes.

We next simulate the GWAS trait. We randomly select $n_{gw}$ (e.g., $50,000$) individuals from UKBB to create the GWAS cohort. We consider the reference panel and GWAS cohort individuals to be non-overlapping. If $p$ is the proportion of non-null genes among $m$ genes, we consider a subset of $m_c=\ceil{mp}$ genes to have a non-zero effect on the trait. Suppose expression data is not available in the GWAS. Assuming that the effect size of local SNPs on expression remains the same between the reference panel and the GWAS populations, we denote the true genetic component of expression of $j^{th}$ non-null gene in GWAS as $G_j=X'_j\bw_j$, where $X_j$ denotes the genotype vector of the local SNPs in GWAS data, and $\bw_j$ denotes the true effect of local SNPs on the expression of $j^{th}$ non-null gene, $j=1,\ldots,m_c$. We simulate $Y=\sum_{j=1}^{m_c} G_j \alpha_j + e$. Assume that $V(Y)=1$ and the total heritability of $Y$ due to the genetic components of expressions of $m_c$ non-null genes is $h^2_y$. Then the random noise $e \sim N(0,1-h^2_y)$. We simulate the effect sizes of the non-null genes independently: $\alpha_j \sim N(0,\frac{h^2_y}{m_c})$, $j=1,\ldots,m_c$. We choose the number of non-null genes per chromosome proportional to the number of locally heritable genes in the chromosome. On each chromosome, we consider the approximate LD blocks identified by Berisa and Pickrell \cite{berisa2016approximately} such that each block contains at least one heritable gene. We consider a subset of LD blocks randomly selected from a given chromosome and randomly choose a gene from each block to be a non-null gene.
In the standard TWAS, we ignore the uncertainty of the predicted genetic component of expression. To evaluate this limitation's impact on {\em polygene}'s performance, we also performed a TWAS using the actual genetic component of expression in the second stage regression based on the GWAS data. We plug in the true effect sizes of the local SNPs on expression while computing the predicted genetic component of expression in the GWAS data. We refer to this approach as the benchmark TWAS. We compare the performance of {\em polygene} applied to the standard and benchmark TWAS statistics. For a given dataset, we estimate the hyperparameters to be used in {\em polygene} based on another dataset generated under the same simulation scenario using the MOM approach discussed above. We used 1000 genome data to estimate the LD structure of SNPs.  

In the reference panel, we choose the local heritability of a gene at random between $10\%-15\%$. We consider $10\%$ of the local SNPs to have a non-zero effect on the expression. We also consider a maximum of 300 SNPs for each gene. While simulating the GWAS trait, we consider a $p$ proportion of locally heritable genes to have a non-null effect on the trait. We chose five different values of $p = 1\%, 2\%, 3\%, 4\%, 5\%$. Thus, for 2988 genes considered on chromosomes 7-22, the maximum number of non-null genes is considered to be 150. We consider two different choices of the sample size of the reference panel $n_E=1000, 4000$, and the sample size of the GWAS data as $n_{GW} = 50000 \text{ } (50K), 70K$ (four different combinations of $n_E$ and $n_{GW}$). We consider two different scenarios of trait heritability due to predicted expressions. In the first scenario, the heritability increases with the proportion of non-null genes, $10 \times x\%$ heritability due to $x\%$ non-null genes, $x=1,2,3,4,5$. In the second scenario, we fix the heritability at $10\%$ and $20\%$. We run the benchmark TWAS for $n_E=1000$ and $n_{GW} = 50K$.

\subsubsection*{Simulation results}

\begin{figure}[H]
\centering

\begin{subfigure}{0.45\textwidth}
  \centering
    \includegraphics[width=1.1\linewidth]{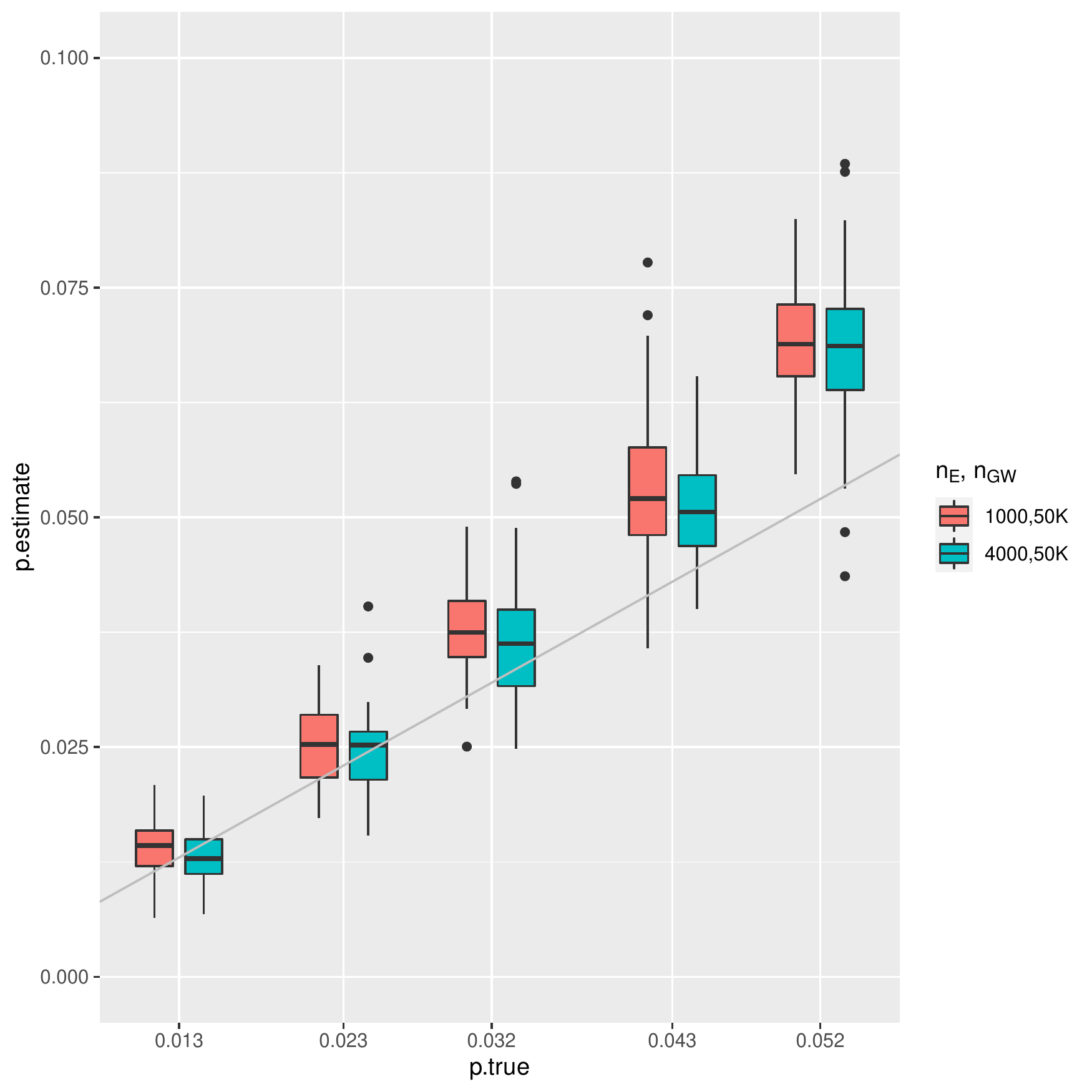}
  \caption{$h^2 = 10\%,20\%,30\%\,40\%,50\%$}
  \label{fig:sub1}
\end{subfigure} \\
\begin{subfigure}{0.45\textwidth}
  \centering
    \includegraphics[width=1\linewidth]{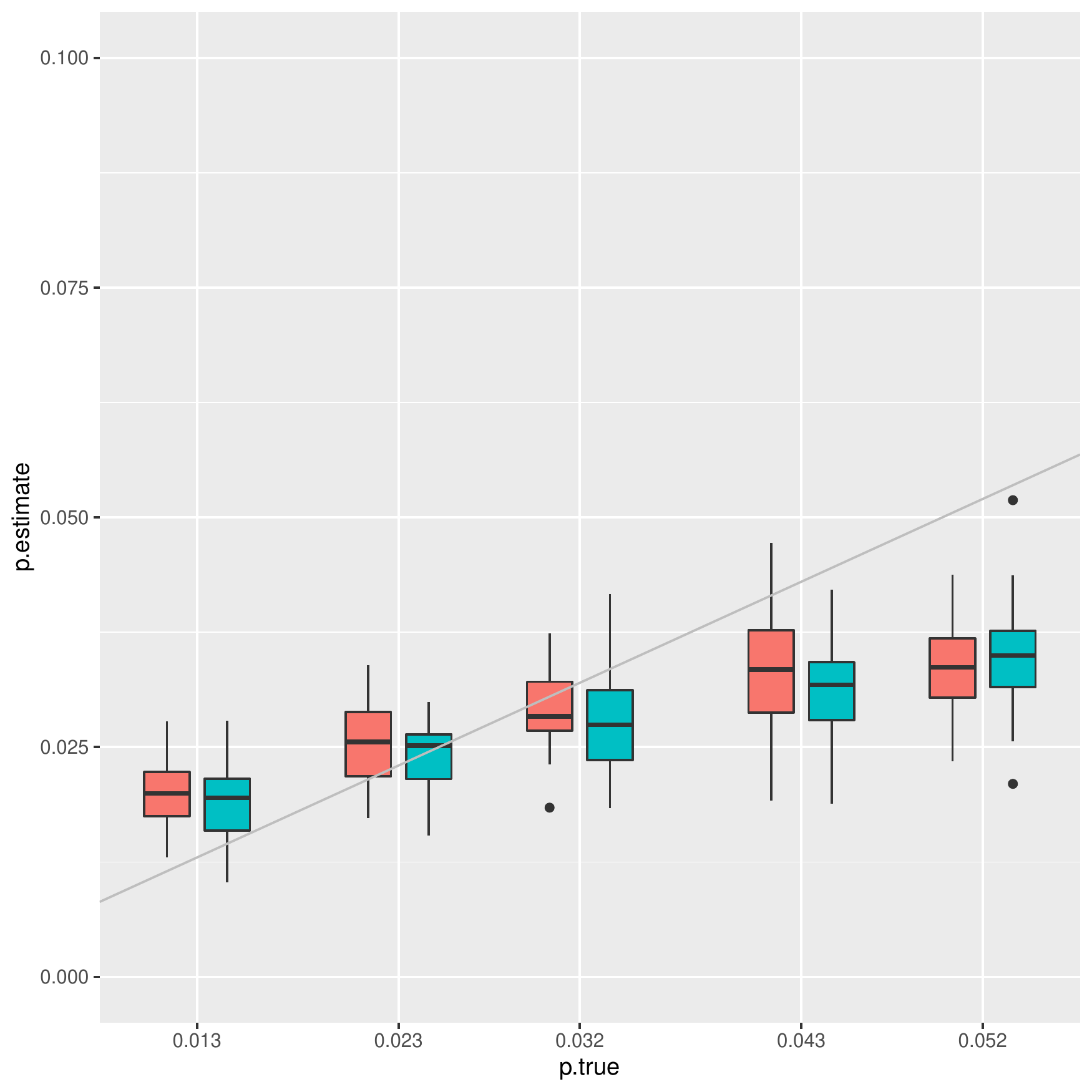}
  \caption{$h^2 = 20\%$}
  \label{fig:sub1}
\end{subfigure}%
\begin{subfigure}{0.45\textwidth}
  \centering
  \includegraphics[width=1.1\linewidth]{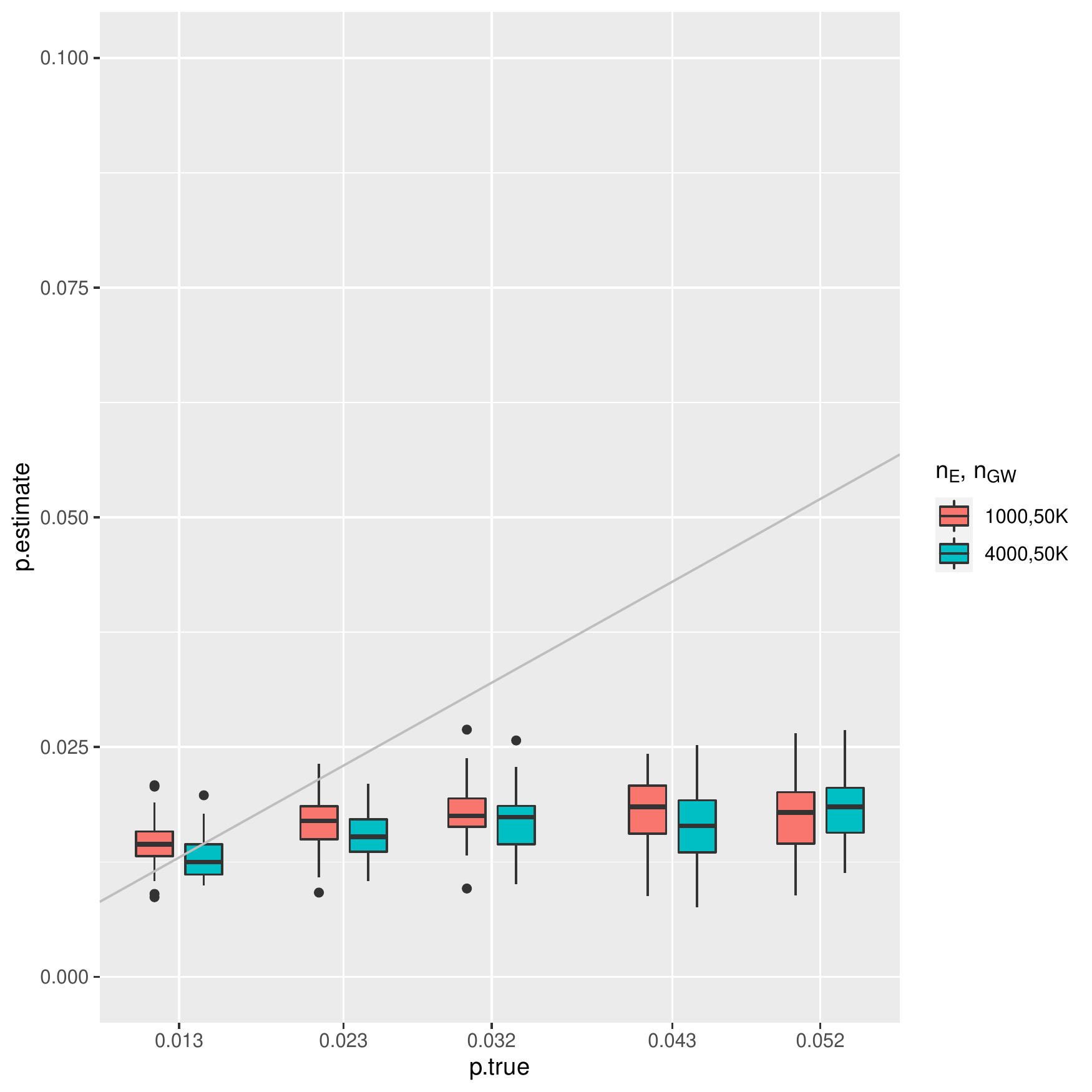}
  \caption{$h^2 = 10\%$}
  \label{fig:sub2}
\end{subfigure}
\caption{Box plots for estimated polygenicity when trait heritability due to the genetic component of expressions are (a) $10\%,20\%,30\%,40\%,50\%$ for $1\%,2\%,3\%,4\%,5\%$ non-null genes, respectively, (b) fixed at $20\%$, (c) fixed at $10\%$. We consider two different choices of the sample size of the expression panel data $(n_E)$ as 1000 and 4000. Here, the GWAS sample size is $50K$.}
\label{fig:est-p-nE-1K-4K-nGW-50K}
\end{figure}

Recall that $n_E$ and $n_{GW}$ denote the sample size of the reference panel expression and GWAS data, respectively; $p$ denotes the true proportion of non-null genes; $h^2$ denotes the trait heritability due to genetic component of expressions. We measure the bias in estimation of $p$ using the relative bias = $\frac{\text{estimated } p \mbox{ } - \mbox{ } \text{true } p}{\text{true } p}$. In simulation scenarios when $h^2$ varies in the range of $10\% - 20\%$ with $n_{GW}=50K$ and $n_E=1000,4000$, {\em polygene} provides an accurate estimates of $p$ when its true values are $1\%, 2\%$ (Fig \ref{fig:est-p-nE-1K-4K-nGW-50K}, Table \ref{tb:relative-bias-p-estimation}). However, when $n_{GW}$ increases to $70K$ in these scenarios, the upward bias increases (Fig \ref{fig:est-p-nE-1K-4K-nGW-50K}, Table \ref{tb:relative-bias-p-estimation}). In such cases, an increase of $n_E$ helps reduce the upward bias. For example, when $p=2\%$ and $h^2=20\%$ with $n_{GW}=70K$, an increase of $n_E$ from $1000$ to $4000$ reduces the mean relative bias from $30\%$ to $24\%$ (Table \ref{tb:relative-bias-p-estimation}). 
For validation of {\em polygene}, we also considered larger values of $h^2$ with $p$ increasing proportionally. {\em polygene} produced substantial upward bias in estimated $p$ when $h^2$ is $30\%,40\%$ and $50\%$ for $p = 3\%, 4\%$ and $5\%$, respectively (Fig \ref{fig:est-p-nE-1K-4K-nGW-50K}). An increase of $n_E$ helped to decrease the upward bias (Fig \ref{fig:est-p-nE-1K-4K-nGW-50K}a, Table \ref{tb:relative-bias-p-estimation}). A possible explanation is that the noise in the predicted expression decreases with an increase of $n_E$. We also note that in realistic scenarios the heritability of a trait due to genetic component of expressions is limited and not expected to be greater than $20\%$ \cite{yao2020quantifying}.
 
\begin{figure}[H]
\centering
\begin{subfigure}{0.45\textwidth}
  \centering
    \includegraphics[width=1\linewidth]{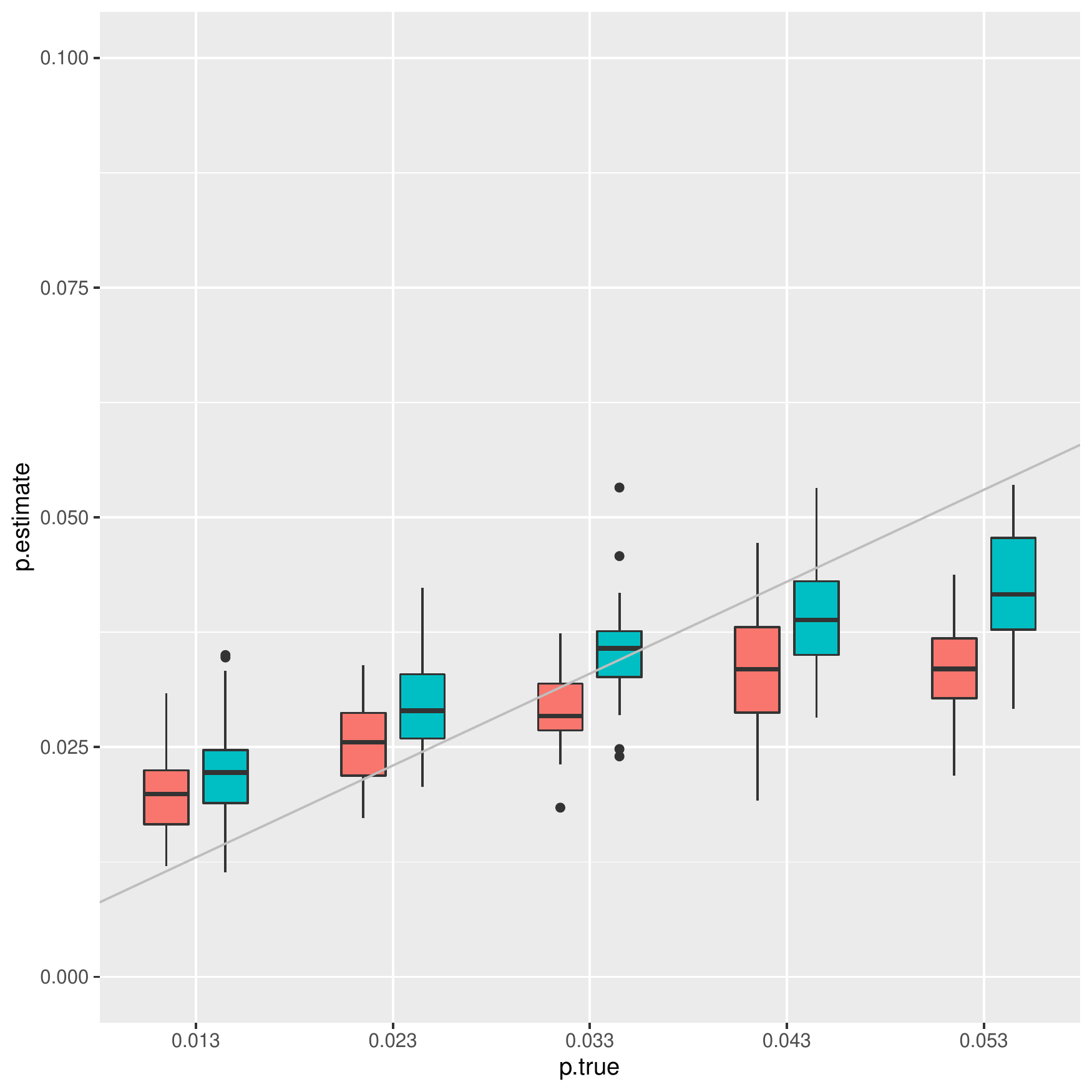}
  \caption{$h^2 = 20\%, n_E=1000$}
  \label{fig:sub1}
\end{subfigure} 
\begin{subfigure}{0.45\textwidth}
  \centering
    \includegraphics[width=1.1\linewidth]{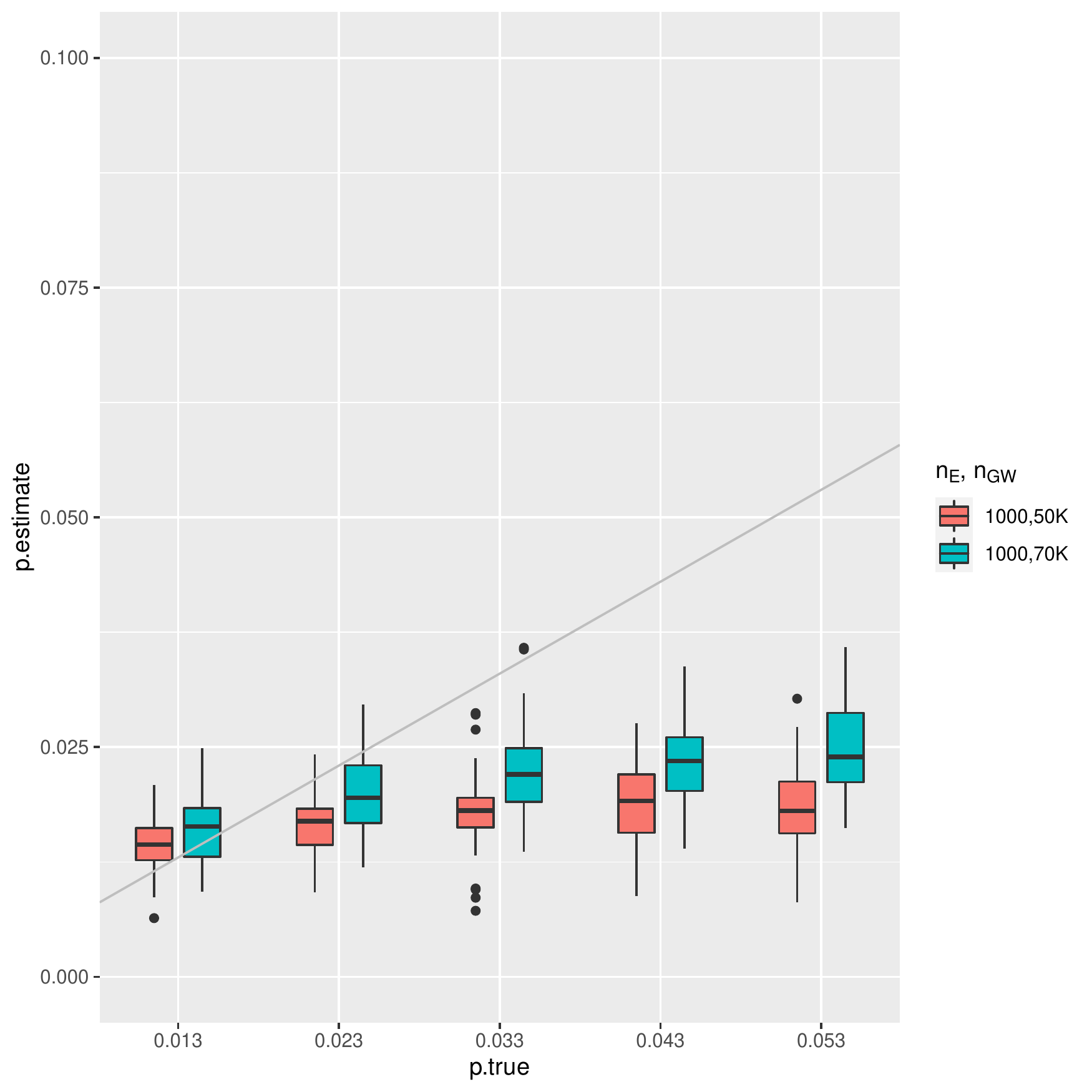}
  \caption{$h^2 = 10\%, n_E=1000$}
  \label{fig:sub1}
\end{subfigure} \\
\begin{subfigure}{0.45\textwidth}
  \centering
    \includegraphics[width=1\linewidth]{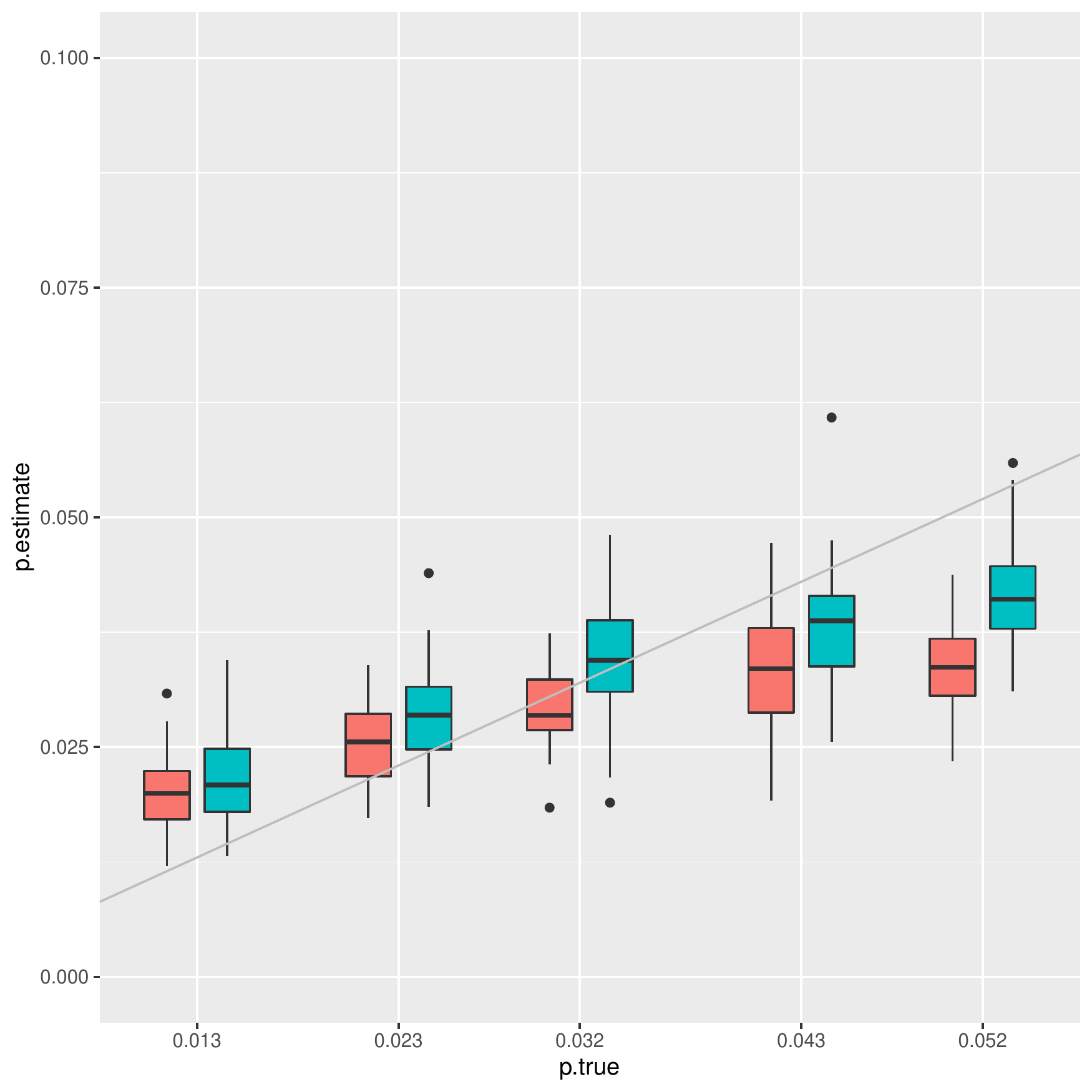}
  \caption{$h^2 = 20\%, n_E=4000$}
  \label{fig:sub1}
\end{subfigure}
\begin{subfigure}{0.45\textwidth}
  \centering
  \includegraphics[width=1.1\linewidth]{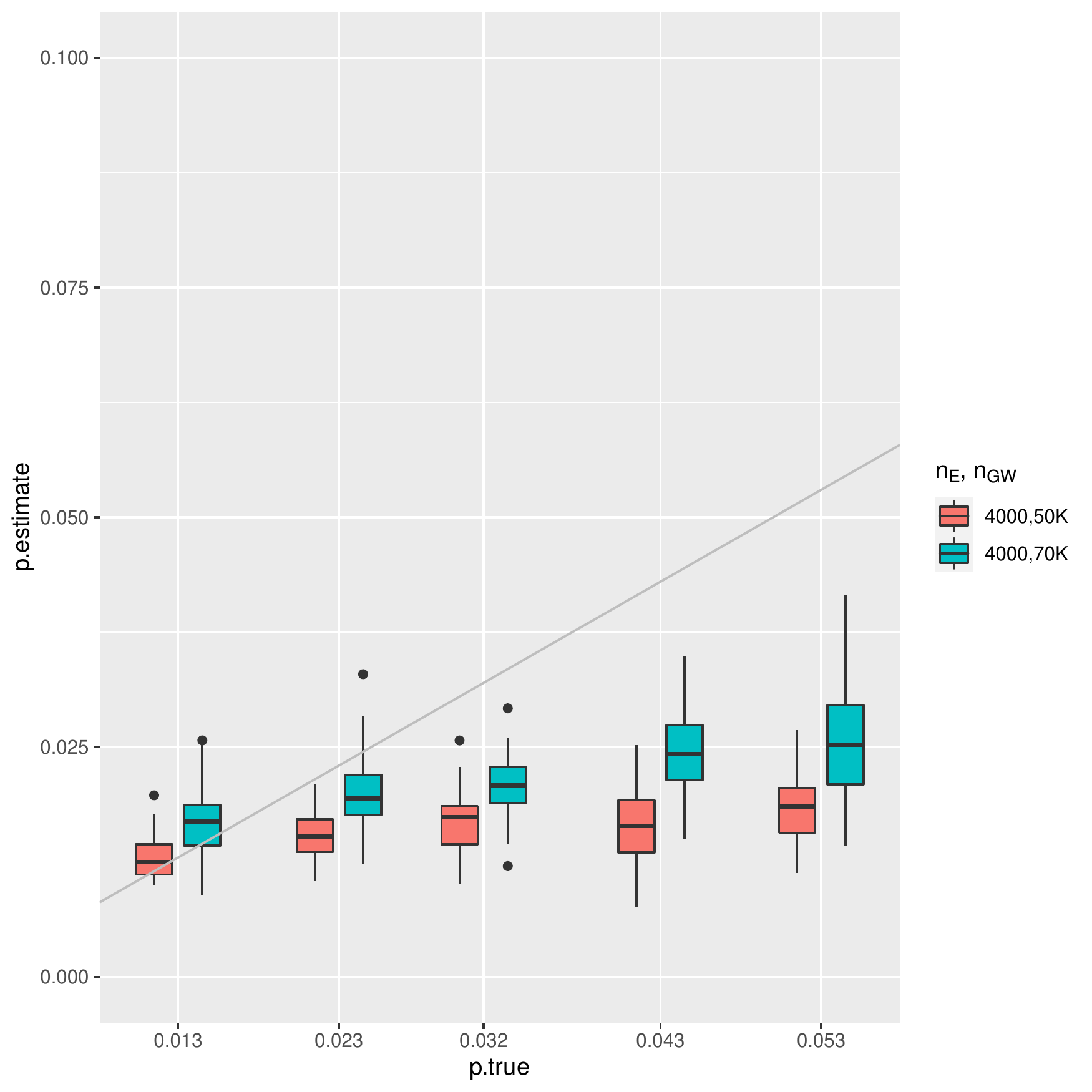}
  \caption{$h^2 = 10\%, n_E=4000$}
  \label{fig:sub2}
\end{subfigure}
\caption{Estimated polygenicity when the GWAS sample size increases from $50K$ to $70K$ for a fixed choice of the sample size of reference panel expression data $(n_E)$. In (a) and (b), we fix $n_E=1000$ and consider the trait heritability due to genetic component of expressions as $20\%$ (a) and $10\%$ (b), respectively. In both scenarios, we increase the GWAS sample size from $50K$ to $70K$. We set $n_E=4000$ and repeat the same analyses in (c) and (d).}
\label{fig:est-p-increase-nGW}
\end{figure}

Next, we discuss the results when $p$ increases for a fixed $h^2$. For $h^2=20\%$ and $p=1\%$, {\em polygene} produces an upward bias (Table \ref{tb:relative-bias-p-estimation}), which may be explained by a larger value of mean $h^2$ per non-null gene. As $p$ increases to $3\%,4\%,5\%$, the mean heritability per non-null gene becomes smaller, and {\em polygene} underestimates $p$ (Fig \ref{fig:est-p-nE-1K-4K-nGW-50K}, Table \ref{tb:relative-bias-p-estimation}). Encouragingly, the downward bias diminishes rapidly with an increase of $n_{GW}$ (Fig \ref{fig:est-p-increase-nGW}). For example, when $h^2=20\%$ and $p=4\%$ with $n_E=1000$, the mean relative bias is $-9\%$ for $n_{GW} = 70K$ compared to $-22\%$ for $n_{GW} = 50K$ (Table \ref{tb:relative-bias-p-estimation}). In overall, the relative downward bias reduced by $12\%-19\%$ with a mean of $15\%$ due to $20K$ increase of $n_{GW}$ ($n_E = 1000$). However, the downward bias did not reduce with an increase of $n_E$ (Fig \ref{fig:est-p-nE-1K-4K-nGW-50K}b, \ref{fig:est-p-nE-1K-4K-nGW-50K}c). The q-value approach \cite{storey2003positive} which only takes p-values as the input and ignores any possible covariance structure of TWAS statistics produced a very large upward bias while estimating $p$ for most of the simulation scenarios (Table \ref{tb:qvalue}). 
 
\begin{figure}[H]
\centering

\begin{subfigure}{0.45\textwidth}
  \centering
    \includegraphics[width=1.1\linewidth]{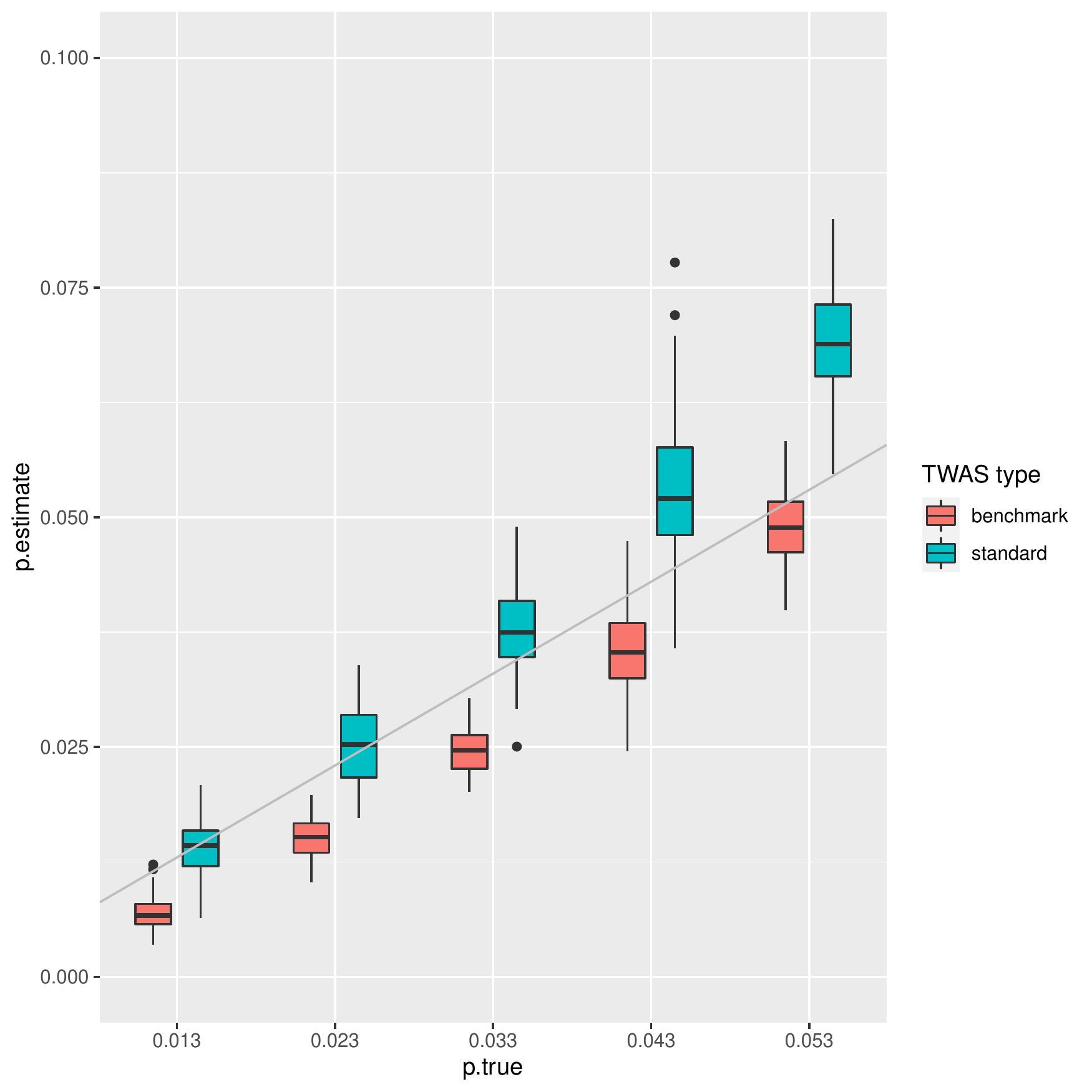}
  \caption{$h^2 = 10\%,20\%,30\%\,40\%,50\%$}
  \label{fig:sub1}
\end{subfigure} \\
\begin{subfigure}{0.45\textwidth}
  \centering
    \includegraphics[width=1\linewidth]{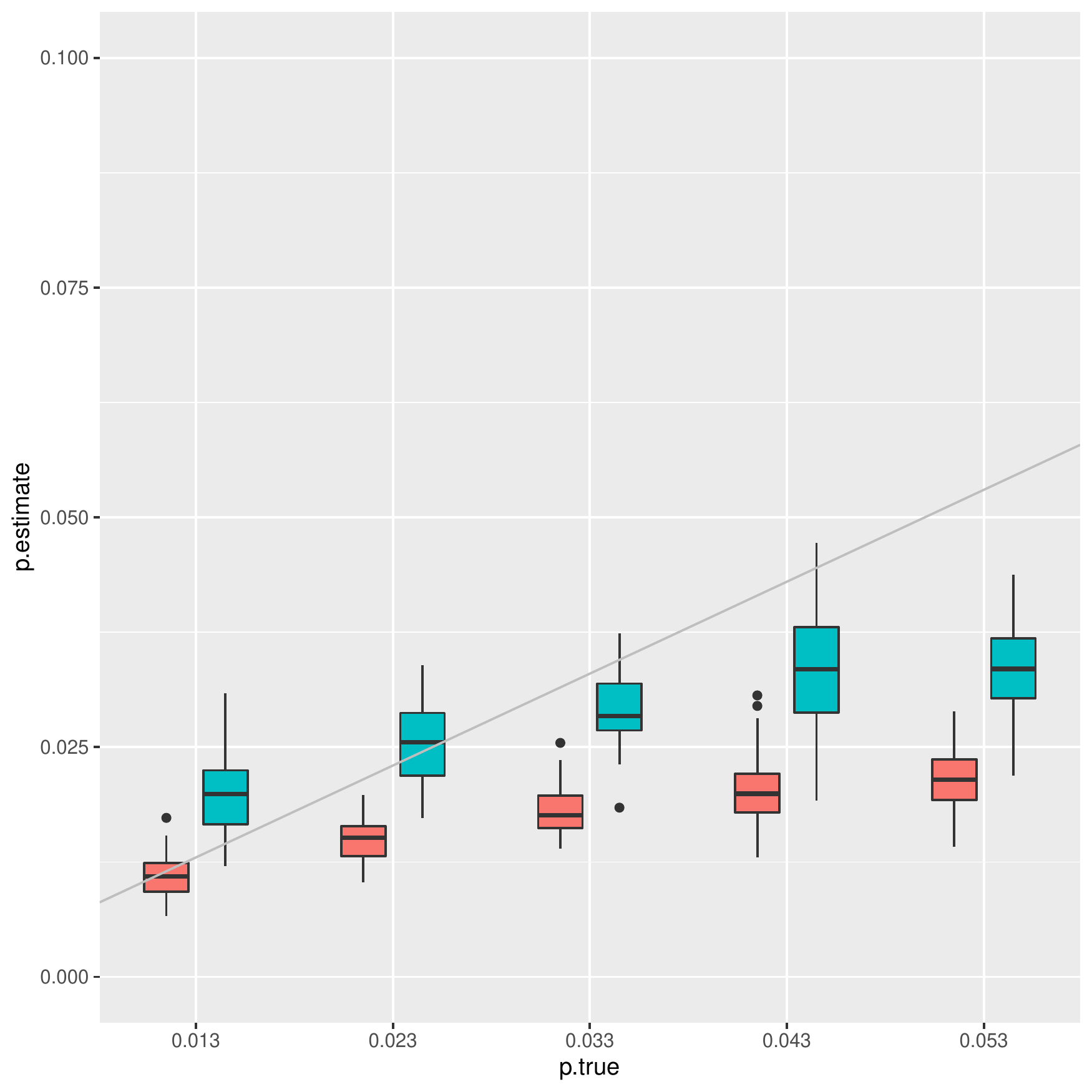}
  \caption{$h^2 = 20\%$}
  \label{fig:sub1}
\end{subfigure}%
\begin{subfigure}{0.45\textwidth}
  \centering
  \includegraphics[width=1.1\linewidth]{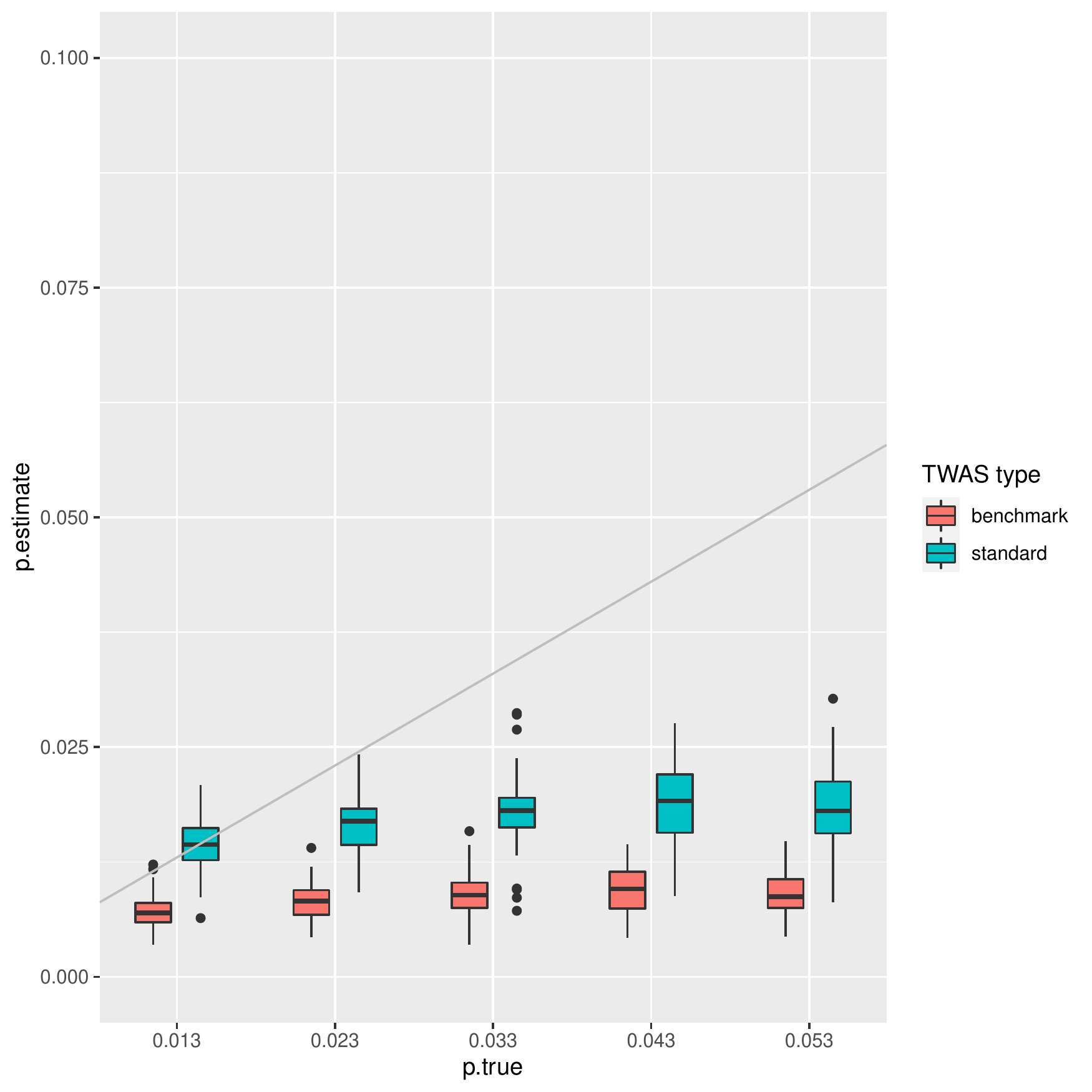}
  \caption{$h^2 = 10\%$}
  \label{fig:sub2}
\end{subfigure}
\caption{Comparison between estimated polygenicity obtained by {\em polygene} using the benchmark TWAS and standard TWAS when the trait heritability due to genetic component of expressions are (a) $10\%,20\%,30\%,40\%,50\%$ for $1\%,2\%,3\%,4\%,5\%$ non-null genes, respectively, (b) fixed at $20\%$, (c) fixed at $10\%$. The sample size of the expression panel and GWAS data are considered as $1000$ and $50K$, respectively.}
\label{fig:benchmark-vs-standard-TWAS}
\end{figure}

In standard TWAS, we ignore the uncertainty of predicted expression. To explore the effect of this limitation, we also ran {\em polygene} for benchmark TWAS in which we plugged in the true effects of local SNPs on the expression in the second stage regression. We observe that {\em polygene} based on the benchmark TWAS consistently underestimates $p$, both in the scenarios of $h^2$ increasing with $p$ and remaining fixed regardless of $p$ (Fig \ref{fig:benchmark-vs-standard-TWAS}a, \ref{fig:benchmark-vs-standard-TWAS}b, \ref{fig:benchmark-vs-standard-TWAS}c, Table \ref{tb:relative-bias-p-estimation}). This indicates that the TWAS framework is underpowered in general. Overall, {\em polygene} estimates $p$ reasonably well, and we can improve the estimation accuracy by increasing the sample sizes of the reference panel expression and GWAS data.

Next, we discuss the usefulness of {\em polygene} while identifying the optimal subset of non-null genes. We measure the selection accuracy by specificity and sensitivity, where specificity measures the proportion of null genes excluded from the inferred subset, and sensitivity measures the proportion of non-null genes included in the subset. {\em polygene} produces a very high level of specificity (mean specificity $\geq 95\%$) consistently across the various simulation scenarios (Table \ref{tb:specificity}). It produces a decent overall sensitivity across the scenarios (Table \ref{tb:sensitivity}). The mean sensitivity is higher when the mean $h^2$ per non-null gene is higher. For $n_E=1000$ and $n_{GW}=50K$, when $h^2$ increases proportionally to $20\%,30\%$ with $p = 2\%,3\%$, the mean sensitivity is $50\%, 55\%$ compared to $33\%,26\%$ when $h^2$ is fixed at $10\%$ (Table \ref{tb:sensitivity}). Thus, the sensitivity decreases with the decrease in the mean heritability per non-null gene. 

Sensitivity increases as $n_E$ and $n_{GW}$ increase. For $n_E=1000$, the mean sensitivity increases when $n_{GW}$ increases from $50K$ to $70K$. For example, for $h^2=10\%$ and $p=2\%$, the mean sensitivity increases from $33\%$ to $40\%$ as $n_{GW}$ increases from $50K$ to $70K$ (Table \ref{tb:sensitivity}). Also, for a given choice of $n_{GW}$, the mean sensitivity increases when $n_E$ increases. For example, for $n_{GW} = 50K$, $h^2=20\%$ and $p=5\%$, mean sensitivity increases from $34\%$ to $39\%$ when $n_E$ increases from 1000 to 4000 (Table \ref{tb:sensitivity}). A joint increase in $n_E$ and $n_{GW}$ provides the maximum sensitivity. For example, for $h^2=10\%$ and $p=2\%$, {\em polygene} produces a maximum of $44\%$ mean sensitivity (Table \ref{tb:sensitivity}) when both the choices of $n_E$ and $n_{GW}$ are largest ($n_E=4000$ and $n_{GW}=70K$). {\em polygene} based on the benchmark TWAS consistently produces a marginally higher $(1-3\%)$ specificity compared to standard TWAS (Table \ref{tb:specificity}) at the expense of a marginally lower sensitivity $(1\%-3\%)$ in some simulation scenarios (Table \ref{tb:sensitivity}). In overall, {\em polygene} performs well to identify the subset of non-null genes with good accuracy. 

\subsection*{Real data application}

We applied {\em polygene} to seven phenotypes in UK Biobank (UKBB), three anthropometric traits, three lipid traits, and a case-control trait. We analyzed chromosomes 1-22 together while integrating expression prediction models available from the software package Fusion \cite{gusev2016integrative} which were fitted for various tissue types in different expression panels. For each trait, we considered a primary and a secondary tissue type. We use the TWAS statistics based on the secondary tissue type to implement the MOM approach to estimate the hyperparameters in the Bayesian framework of {\em polygene}. We finally implement {\em polygene} based on the trait's TWAS data obtained from the primary tissue type. Finucane et al. \cite{finucane2018heritability} developed a novel method to identify the relevant tissue or cell types for a complex trait and reported the most appropriate tissue types for a collection of traits. Their findings guide the choice of the tissue types pertinent to the traits considered in our analysis. For example, while analyzing a lipid trait (LDL, HDL, or triglyceride), we considered the liver tissue primary and whole blood secondary tissue type.
Similarly, for WHR, we considered adipose as the primary tissue and muscle-skeletal as the secondary tissue type. We note that Finucane et al. \cite{finucane2018heritability} reported both of these tissue types as significantly relevant for WHR (\cite{finucane2018heritability}). We considered the GTEx expression data for most traits, except whole blood in Young Finish Study (YFS) for asthma as the primary tissue type.

In our analysis, height appeared to be most polygenic (posterior median of $p$ as $23\%$ with the $95\%$ central posterior interval as $(21\%,25\%)$) (Table \ref{polygene-UKBB}). For the other two anthropometric traits, BMI is $11\%$ polygenic with a posterior interval of $9\%-12\%$, and WHR is $7\%$ polygenic with a posterior interval of $6\%-8\%$ (Table \ref{polygene-UKBB}). Among the lipids, HDL is the most polygenic $(10\%)$, and LDL is the least polygenic $(2\%)$ with triglycerides falling between $(7\%)$. The point estimates and the posterior intervals of $p$ did not overlap between HDL and LDL, as well as triglycerides and LDL (Table \ref{polygene-UKBB}). This suggests that HDL and triglycerides are more polygenic than LDL. In our analysis, the case-control trait asthma turned out to be the least polygenic $(0.2\%)$.
 
For each trait analyzed, we also report the optimal subset of non-null genes identified by {\em polygene} (Table \ref{tb:subset-asthma} - \ref{tb:subset-height3}). While the subset for height contains 300 genes (Table \ref{tb:subset-height1}, \ref{tb:subset-height2}, \ref{tb:subset-height3}), the subset for asthma contains five genes (Table \ref{tb:subset-asthma}). Many of the genes included in the optimal subset for a trait were previously reported to be associated with the trait or other relevant traits. For each trait, we mention such a gene in the following. 
For asthma, QSOX1 on chromosome 1 (Table \ref{tb:subset-asthma}) has been reported to be associated with blood protein measurements \cite{sun2018genomic}. For lipid traits, CELSR2 on chromosome 1 (Table \ref{tb:subset-LDL}, \ref{tb:subset-HDL}, \ref{tb:subset-triglycerides}) is previously known to be associated with lipids (numerous studies reported in EBI GWAS catalog). For BMI, PPP2R3A on chromosome 3 (Table \ref{tb:subset-BMI}) has been reported to be associated with BMI \cite{zhu2020shared} in the EBI GWAS catalog. For WHR, GRK4 on chromosome 4 (Table \ref{tb:subset-WHR}) is previously known to be associated with BMI-adjusted hip circumference \cite{justice2021genome}. For height, PEX1 on chromosome 7 (Table \ref{tb:subset-height2}) is known to be associated with height \cite{tachmazidou2017whole} and other anthropometric traits. 

\section*{Discussion}

We propose a Bayesian approach {\em polygene} to estimate the gene-level polygenicity for a complex trait under the framework of TWAS. {\em polygene} simultaneously provides an optimal subset of non-null genes for the trait. It explicitly accounts for the covariance structure between the genes. The method uses summary-level TWAS association statistics which are often publicly available. Along with the point estimate of gene-level polygenicity, {\em polygene} provides a posterior interval to assess the uncertainty. 

The simulation study shows that {\em polygene} performs well in estimating the proportion of non-null genes in realistic scenarios. When it produces an upward bias (for large values of trait heritability), increasing the sample size of the reference panel expression data helps reduce the bias. The downward bias produced by {\em polygene} when the heritability per non-null gene is small diminishes rapidly with the increase of the GWAS sample size. For computational convenience, we experimented with a maximum GWAS sample size of $70K$ in simulations. The relative downward bias decreased by an average of 15\% due to a GWAS sample size increase of $20K$. With the sample size of contemporary GWASs available (e.g., hundreds of thousands in UK Biobank), we anticipate that the downward bias will reduce substantially.  In simulations, we considered the choices of $p$ as $1\% - 5\%$. If we increase $p$ for a fixed $h^2$, the magnitude of downward bias will increase due to a decrease in the mean trait heritability per non-null gene. For $h^2$ proportionally increasing with $p$, {\em polygene} is expected to perform well. Consistent downward bias in estimated gene-level polygenicity observed using the benchmark TWAS indicates that the TWAS framework may be underpowered. It also suggests that some portion of the upward bias produced by {\em polygene} while using the standard TWAS may be due to the lack of uncertainty adjustment in the predicted expression. 

{\em polygene} performs well while identifying the optimal subset of non-null genes. The estimated subsets are highly specific, with an overall good level of sensitivity. The sensitivity improves if the sample size of the reference panel expression and GWAS data increase. Simulations show that the selection accuracy of {\em polygene} is comparable between standard and benchmark TWAS. Thus, functional studies can further investigate the optimal subset of non-null genes identified for a complex trait.

While developing the Bayesian method, we also experimented with the Dirac spike and slab prior (point mass at zero). Our preliminary simulations showed that the MCMC implementation does not perform robustly, particularly for many genes (in thousands). We also found that the MCMC for the continuous spike and slab prior is computationally faster than that for the Dirac spike and slab prior. We implemented a method of moments (MOM) approach to choose the hyperparameters in the model based on a tissue type closely relevant to the trait. We propose using the TWAS statistics from a tissue type nearly appropriate to the trait. If the data for a secondary tissue type is unavailable, we can always apply the MOM approach for the primary tissue (an entirely empirical Bayes approach). We have described {\em polygene} for a continuous trait. However, the method can be applied to a binary trait as well, mainly because the form of the covariance matrix for marginal TWAS statistics depends on the expression prediction model obtained from the reference panel and the LD matrix of the eQTLs. Focus \cite{mancuso2019probabilistic} is a fine-mapping method under the TWAS framework which also considers covariance between the TWAS statistics. We note that the analytical form of covariance is similar between Focus and {\em polygene}. In this paper, we use the term `non-null gene' instead of a `causal gene'. The genetic component of the expression of a non-null gene has a non-zero effect on the trait. Our definition does not implicate a formally causal biological relationship between the non-null gene and the trait. In future work, we plan to estimate the gene-level polygenicity while adjusting for the uncertainty of predicted expression in TWAS.

In summary, {\em polygene} is a methodologically sound unified Bayesian approach to estimating the gene-level polygenicity and optimal subset of non-null genes for a complex trait under the TWAS framework. It is computationally efficient and can be applied to various traits to understand their genetic architecture better.
  
\section*{Data availability} 
The datasets that we have analyzed in this paper are available (either openly or via applications) from the following websites:\\
UK Biobank: \url{https://www.ukbiobank.ac.uk/} \\
GTEx: \url{https://gtexportal.org/home/} \\
Fusion: \url{http://gusevlab.org/projects/fusion/}

\section*{Acknowledgments}

We acknowledge Dr. Nicholas Mancuso for helpful discussions related to this work. We sincerely thank Dr. Tanushree Haldar for helping with the graphical presentation of the paper. This research was conducted using the UK Biobank resource under applications 33297 and 33127.

\bibliography{reference.bib}

\begin{thebibliography}{10}

\bibitem{zhang2018estimation}
Yan Zhang, Guanghao Qi, Ju-Hyun Park, and Nilanjan Chatterjee.
\newblock Estimation of complex effect-size distributions using summary-level
  statistics from genome-wide association studies across 32 complex traits.
\newblock {\em Nature Genetics}, 50(9):1318--1326, 2018.

\bibitem{chatterjee2013projecting}
Nilanjan Chatterjee, Bill Wheeler, Joshua Sampson, Patricia Hartge, Stephen~J
  Chanock, and Ju-Hyun Park.
\newblock Projecting the performance of risk prediction based on polygenic
  analyses of genome-wide association studies.
\newblock {\em Nature Genetics}, 45(4):400--405, 2013.

\bibitem{o2019extreme}
Luke~J O'Connor, Armin~P Schoech, Farhad Hormozdiari, Steven Gazal, Nick
  Patterson, and Alkes~L Price.
\newblock Extreme polygenicity of complex traits is explained by negative
  selection.
\newblock {\em The American Journal of Human Genetics}, 105(3):456--476, 2019.

\bibitem{johnson2021estimation}
Ruth Johnson, Kathryn~S Burch, Kangcheng Hou, Mario Paciuc, Bogdan Pasaniuc,
  and Sriram Sankararaman.
\newblock Estimation of regional polygenicity from gwas provides insights into
  the genetic architecture of complex traits.
\newblock {\em PLoS computational biology}, 17(10):e1009483, 2021.

\bibitem{gamazon2015gene}
Eric~R Gamazon, Heather~E Wheeler, Kaanan~P Shah, Sahar~V Mozaffari, Keston
  Aquino-Michaels, Robert~J Carroll, Anne~E Eyler, Joshua~C Denny, Dan~L
  Nicolae, Nancy~J Cox, et~al.
\newblock A gene-based association method for mapping traits using reference
  transcriptome data.
\newblock {\em Nature genetics}, 47(9):1091, 2015.

\bibitem{gusev2016integrative}
Alexander Gusev, Arthur Ko, Huwenbo Shi, Gaurav Bhatia, Wonil Chung, Brenda~WJH
  Penninx, Rick Jansen, Eco~JC De~Geus, Dorret~I Boomsma, Fred~A Wright, et~al.
\newblock Integrative approaches for large-scale transcriptome-wide association
  studies.
\newblock {\em Nature genetics}, 48(3):245, 2016.

\bibitem{mancuso2018large}
Nicholas Mancuso, Simon Gayther, Alexander Gusev, Wei Zheng, Kathryn~L Penney,
  Zsofia Kote-Jarai, Rosalind Eeles, Matthew Freedman, Christopher Haiman, and
  Bogdan Pasaniuc.
\newblock Large-scale transcriptome-wide association study identifies new
  prostate cancer risk regions.
\newblock {\em Nature communications}, 9(1):1--11, 2018.

\bibitem{siewert2022leveraging}
Katherine~M Siewert-Rocks, Samuel~S Kim, Douglas~W Yao, Huwenbo Shi, and
  Alkes~L Price.
\newblock Leveraging gene co-regulation to identify gene sets enriched for
  disease heritability.
\newblock {\em The American Journal of Human Genetics}, 109(3):393--404, 2022.

\bibitem{george1993variable}
Edward~I George and Robert~E McCulloch.
\newblock Variable selection via gibbs sampling.
\newblock {\em Journal of the American Statistical Association},
  88(423):881--889, 1993.

\bibitem{ishwaran2005spike}
Hemant Ishwaran and J~Sunil Rao.
\newblock Spike and slab variable selection: frequentist and bayesian
  strategies.
\newblock {\em The Annals of Statistics}, 33(2):730--773, 2005.

\bibitem{rovckova2018bayesian}
Veronika Ro{\v{c}}kov{\'a}.
\newblock Bayesian estimation of sparse signals with a continuous
  spike-and-slab prior.
\newblock {\em The Annals of Statistics}, 46(1):401--437, 2018.

\bibitem{malsiner2018comparing}
Gertraud Malsiner-Walli and Helga Wagner.
\newblock Comparing spike and slab priors for bayesian variable selection.
\newblock {\em arXiv preprint arXiv:1812.07259}, 2018.

\bibitem{tibshirani1996regression}
Robert Tibshirani.
\newblock Regression shrinkage and selection via the lasso.
\newblock {\em Journal of the Royal Statistical Society: Series B
  (Methodological)}, 58(1):267--288, 1996.

\bibitem{zou2005regularization}
Hui Zou and Trevor Hastie.
\newblock Regularization and variable selection via the elastic net.
\newblock {\em Journal of the royal statistical society: series B (statistical
  methodology)}, 67(2):301--320, 2005.

\bibitem{gtex2015genotype}
GTEx Consortium et~al.
\newblock The genotype-tissue expression (gtex) pilot analysis: multitissue
  gene regulation in humans.
\newblock {\em Science}, 348(6235):648--660, 2015.

\bibitem{gtex2017genetic}
GTEx Consortium et~al.
\newblock Genetic effects on gene expression across human tissues.
\newblock {\em Nature}, 550(7675):204, 2017.

\bibitem{berisa2016approximately}
Tomaz Berisa and Joseph~K Pickrell.
\newblock Approximately independent linkage disequilibrium blocks in human
  populations.
\newblock {\em Bioinformatics}, 32(2):283, 2016.

\bibitem{yao2020quantifying}
Douglas~W Yao, Luke~J O?connor, Alkes~L Price, and Alexander Gusev.
\newblock Quantifying genetic effects on disease mediated by assayed gene
  expression levels.
\newblock {\em Nature genetics}, 52(6):626--633, 2020.

\bibitem{storey2003positive}
John~D Storey.
\newblock The positive false discovery rate: a bayesian interpretation and the
  q-value.
\newblock {\em The annals of statistics}, 31(6):2013--2035, 2003.

\bibitem{finucane2018heritability}
Hilary~K Finucane, Yakir~A Reshef, Verneri Anttila, Kamil Slowikowski,
  Alexander Gusev, Andrea Byrnes, Steven Gazal, Po-Ru Loh, Caleb Lareau, Noam
  Shoresh, et~al.
\newblock Heritability enrichment of specifically expressed genes identifies
  disease-relevant tissues and cell types.
\newblock {\em Nature genetics}, 50(4):621, 2018.

\bibitem{sun2018genomic}
Benjamin~B Sun, Joseph~C Maranville, James~E Peters, David Stacey, James~R
  Staley, James Blackshaw, Stephen Burgess, Tao Jiang, Ellie Paige, Praveen
  Surendran, et~al.
\newblock Genomic atlas of the human plasma proteome.
\newblock {\em Nature}, 558(7708):73--79, 2018.

\bibitem{zhu2020shared}
Zhaozhong Zhu, Yanjun Guo, Huwenbo Shi, Cong-Lin Liu, Ronald~Allan Panganiban,
  Wonil Chung, Luke~J O'Connor, Blanca~E Himes, Steven Gazal, Kohei Hasegawa,
  et~al.
\newblock Shared genetic and experimental links between obesity-related traits
  and asthma subtypes in uk biobank.
\newblock {\em Journal of Allergy and Clinical Immunology}, 145(2):537--549,
  2020.

\bibitem{justice2021genome}
Anne~E Justice, Kristin Young, Stephanie~M Gogarten, Tamar Sofer, Misa Graff,
  Shelly Ann~M Love, Yujie Wang, Yann~C Klimentidis, Miguel Cruz, Xiuqing Guo,
  et~al.
\newblock Genome-wide association study of body fat distribution traits in
  hispanics/latinos from the hchs/sol.
\newblock {\em Human molecular genetics}, 30(22):2190--2204, 2021.

\bibitem{tachmazidou2017whole}
Ioanna Tachmazidou, D{\'a}niel S{\"u}veges, Josine~L Min, Graham~RS Ritchie,
  Julia Steinberg, Klaudia Walter, Valentina Iotchkova, Jeremy
  Schwartzentruber, Jie Huang, Yasin Memari, et~al.
\newblock Whole-genome sequencing coupled to imputation discovers genetic
  signals for anthropometric traits.
\newblock {\em The American Journal of Human Genetics}, 100(6):865--884, 2017.

\bibitem{mancuso2019probabilistic}
Nicholas Mancuso, Malika~K Freund, Ruth Johnson, Huwenbo Shi, Gleb Kichaev,
  Alexander Gusev, and Bogdan Pasaniuc.
\newblock Probabilistic fine-mapping of transcriptome-wide association studies.
\newblock {\em Nature genetics}, 51(4):675--682, 2019.

\end{thebibliography}
\bibliographystyle{unsrt}

\clearpage

\clearpage

\clearpage

\clearpage

\begin{table}[]
\centering
\caption{Percentage of relative bias produced by {\em polygene} while estimating $p$ using standard TWAS and benchmark TWAS in various simulation scenarios. $n_E$ and $n_{GW}$ denote the sample sizes of the reference panel expression and the GWAS data, respectively.}
\begin{tabular}{ccccccc}
\hline
\multicolumn{2}{c}{Simulation} & \multicolumn{4}{c|}{standard TWAS}                                    & benchmark TWAS  \\ \cline{3-7} 
\multicolumn{2}{c}{scenarios}  & \multicolumn{4}{c}{$n_E,n_{GW}$}                                            & $n_E, n_{GW}$          \\ \hline
$p$             & $h^2$             & $1000,50K$ & $1000,70K$ & $4000,50K$ & $4000,70K$ & $1000,50K$ \\  \hline
1\% & 10\% & 3   & 25  & 4   & 21  & -43 \\
2\% & 20\% & 6   & 30  & 7   & 24  & -33 \\
3\% & 30\% & 14  & 36  & 13  & 37  & -24 \\
4\% & 40\% & 24  & 41  & 19  & 37  & -12 \\
5\% & 50\% & 31  & 53  & 32  & 49  & -6  \\
1\% & 20\% & 47  & 71  & 48  & 67  & -10 \\
3\% & 20\% & -12 & 7   & -12 & 7   & -43 \\
4\% & 20\% & -22 & -9  & -26 & -12 & -51 \\
5\% & 20\% & -37 & -20 & -33 & -20 & -59 \\
2\% & 10\% & -32 & -13 & -31 & -16 & -61 \\
3\% & 10\% & -47 & -32 & -47 & -32 & -70 \\
4\% & 10\% & -57 & -45 & -57 & -44 & -75 \\
5\% & 10\% & -65 & -53 & -63 & -52 & -80 \\ \hline
\end{tabular}
\label{tb:relative-bias-p-estimation}
\end{table}

\clearpage

\begin{table}[]
\centering
\caption{Mean specificity of {\em polygene} while inferring the subset of non-null genes using standard and benchmark TWAS in various simulation scenarios. $n_E$ and $n_{GW}$ denote the sample sizes of the reference panel expression and the GWAS data, respectively.}
\begin{tabular}{ccccccc}
\hline
\multicolumn{2}{c}{Simulation} & \multicolumn{4}{c|}{standard TWAS}                                    & benchmark TWAS  \\ \cline{3-7} 
\multicolumn{2}{c}{scenarios}  & \multicolumn{4}{c}{$n_E,n_{GW}$}                                            & $n_E, n_{GW}$          \\ \hline
$p$             & $h^2$             & $1000,50K$ & $1000,70K$ & $4000,50K$ & $4000,70K$ & $1000,50K$ \\  \hline
1\% & 10\% & 99 & 99 & 99 & 99 & 100 \\
2\% & 20\% & 99 & 98 & 99 & 99 & 100 \\
3\% & 30\% & 98 & 97 & 98 & 98 & 99  \\
4\% & 40\% & 97 & 97 & 97 & 97 & 99  \\
5\% & 50\% & 96 & 95 & 96 & 96 & 98  \\
1\% & 20\% & 99 & 99 & 99 & 99 & 100 \\
3\% & 20\% & 99 & 98 & 99 & 98 & 100 \\
4\% & 20\% & 98 & 98 & 99 & 98 & 99  \\
5\% & 20\% & 98 & 98 & 98 & 98 & 100 \\
2\% & 10\% & 99 & 99 & 99 & 99 & 100 \\
3\% & 10\% & 99 & 99 & 99 & 99 & 100 \\
4\% & 10\% & 99 & 99 & 99 & 99 & 100 \\
5\% & 10\% & 99 & 99 & 99 & 99 & 100 \\ \hline
\end{tabular}
\label{tb:specificity}
\end{table}

\clearpage

\begin{table}[]
\centering
\caption{Mean sensitivity of {\em polygene} while inferring the subset of non-null genes using standard and benchmark TWAS in various simulation scenarios. $n_E$ and $n_{GW}$ denote the sample sizes of the reference panel expression and GWAS data, respectively.}
\begin{tabular}{ccccccc}
\hline
\multicolumn{2}{c}{Simulation} & \multicolumn{4}{c|}{standard TWAS}                                    & benchmark TWAS  \\ \cline{3-7} 
\multicolumn{2}{c}{scenarios}  & \multicolumn{4}{c}{$n_E,n_{GW}$}                                            & $n_E, n_{GW}$          \\ \hline
$p$             & $h^2$             & $1000,50K$ & $1000,70K$ & $4000,50K$ & $4000,70K$ & $1000,50K$ \\  \hline
1\% & 10\% & 44 & 51 & 49 & 55 & 43 \\
2\% & 20\% & 50 & 57 & 53 & 60 & 49 \\
3\% & 30\% & 55 & 61 & 58 & 64 & 53 \\
4\% & 40\% & 58 & 64 & 61 & 67 & 58 \\
5\% & 50\% & 62 & 67 & 65 & 70 & 61 \\
1\% & 20\% & 61 & 65 & 66 & 69 & 60 \\
3\% & 20\% & 44 & 52 & 47 & 54 & 42 \\
4\% & 20\% & 40 & 46 & 42 & 49 & 37 \\
5\% & 20\% & 34 & 42 & 39 & 45 & 32 \\
2\% & 10\% & 33 & 40 & 36 & 44 & 30 \\
3\% & 10\% & 26 & 34 & 30 & 36 & 24 \\
4\% & 10\% & 22 & 27 & 25 & 32 & 19 \\
5\% & 10\% & 18 & 24 & 21 & 28 & 16 \\ \hline
\end{tabular}
\label{tb:sensitivity}
\end{table}

\clearpage

\begin{table}[]
\centering
\caption{Estimation of gene-level polygenicity by {\em polygene} for seven phenotypes in UK Biobank integrating external expression panel data. We used the expression prediction models from the Fusion software package. The expression prediction model based on whole blood, considered a primary tissue type for asthma, was collected from the Young Finish Study (YFS). All other expression prediction models used here were fitted in the GTEx study.}
\label{polygene-UKBB}
\begin{tabular}{cccccc}
\hline
              & \multicolumn{3}{c}{Estimated polygenicity}                          &                 \multicolumn{2}{c}{Tissue-type}                     \\ \cline{2-6}
              & \multicolumn{1}{c|}{Posterior median}   & \multicolumn{2}{c}{95\% posterior interval} &       Primary & Secondary                \\ \cline{5-6}  \hline   
Height        & 22.7\% & 20.9\% & 24.6\% & Muscle skeletal        & Adipose subcutaneous   \\
BMI           & 10.5\% & 8.8\%  & 12.3\% & Brain cerebellum       & Adipose subcutaneous   \\
WHR           & 6.8\%  & 5.8\%  & 7.9\%  & Adipose subcutaneous   & Muscle skeletal        \\
HDL           & 9.9\%  & 7.4\%  & 12.7\% & Liver                  & Whole blood            \\
Triglycerides & 7.0\%  & 5.0\%  & 9.4\%  & Liver                  & Whole blood            \\
LDL           & 1.7\%  & 0.8\%  & 3.1\%  & Liver                  & Whole blood            \\
Asthma        & 0.2\%  & 0.1\%  & 0.4\%  & Whole blood (YFS)           & Whole blood      \\  \hline    
\end{tabular}
\end{table}

\clearpage

\newcommand{\beginsupplement}{%
        \setcounter{table}{0}
        \renewcommand{\thetable}{S\arabic{table}}%
        \setcounter{figure}{0}
        \renewcommand{\thefigure}{S\arabic{figure}}%
        \setcounter{section}{0}
        \renewcommand{\thesection}{\arabic{section}}%
        \setcounter{algorithm}{0}
        \renewcommand{\thealgorithm}{S\arabic{algorithm}}%
}

\beginsupplement

\newcommand{\listXname}{List of Xs}

\pagenumbering{arabic}
\setcounter{page}{1}

\title{Supplementary materials: A Bayesian method for estimating gene-level polygenicity under the framework of transcriptome-wide association study}

\maketitle

\newpage

\clearpage

\begin{figure}[H]
\centering
\begin{subfigure}{0.45\textwidth}
  \centering
    \includegraphics[width=1.1\linewidth]{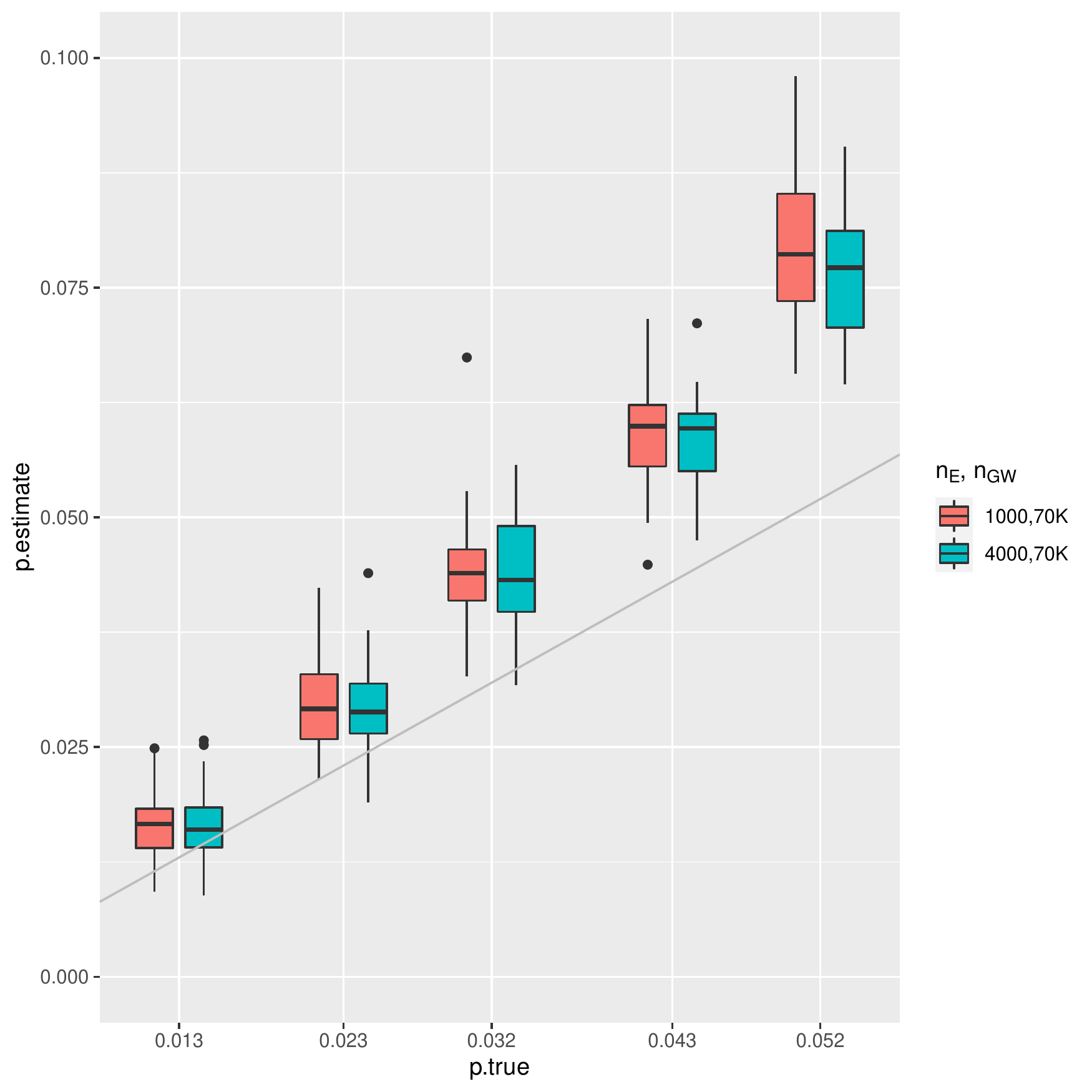}
  \caption{$h^2 = 10\%,20\%,30\%\,40\%,50\%$}
  \label{fig:sub1}
\end{subfigure} \\
\begin{subfigure}{0.45\textwidth}
  \centering
    \includegraphics[width=1\linewidth]{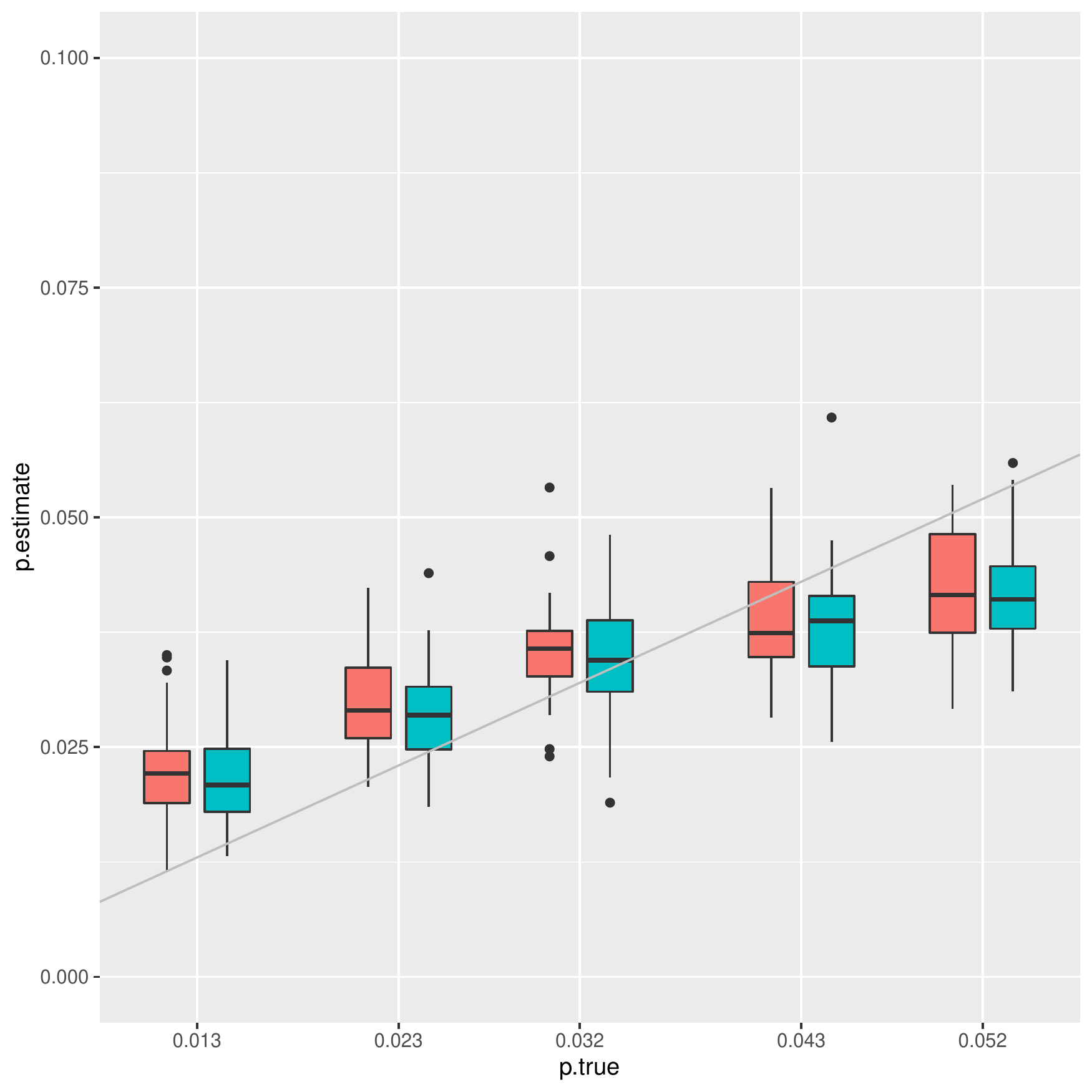}
  \caption{$h^2 = 20\%$}
  \label{fig:sub1}
\end{subfigure}%
\begin{subfigure}{0.45\textwidth}
  \centering
  \includegraphics[width=1.1\linewidth]{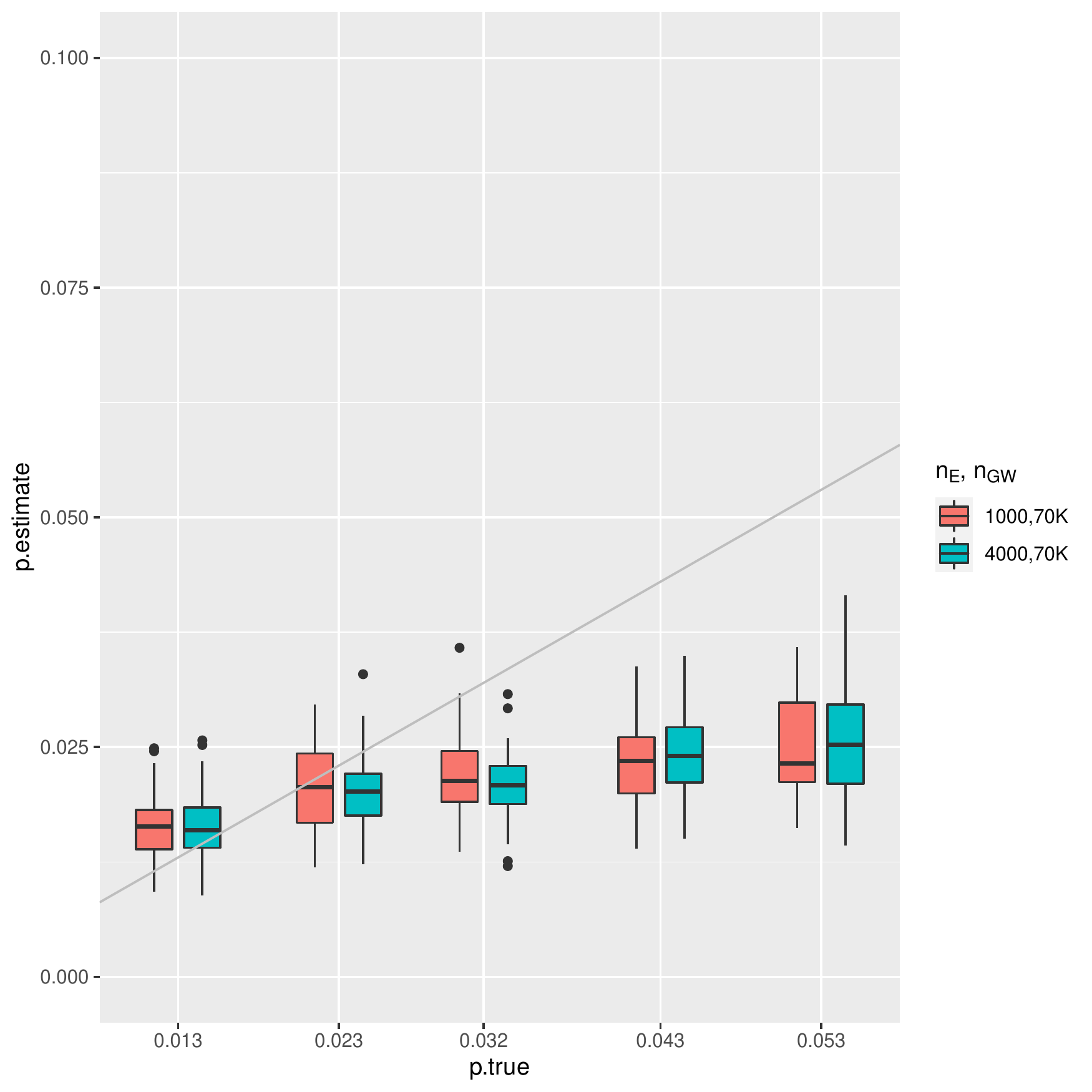}
  \caption{$h^2 = 10\%$}
  \label{fig:sub2}
\end{subfigure}
\caption{Estimated polygenicity for GWAS sample size $70K$. The trait heritability due to the genetic component of expressions are $10\%,20\%,30\%,40\%,50\%$ for $1\%,2\%,3\%,4\%,5\%$ non-null genes, respectively (a), fixed at $20\%$ (b), fixed at $10\%$ (c). We consider two different choices of the sample size of the expression panel data $(n_E)$ as 1000 and 4000.}
\label{fig:supple-est-p-nE-1K-4K-nGW-70K}
\end{figure}

\clearpage

\begin{figure}[H]
\centering

\begin{subfigure}{0.45\textwidth}
  \centering
    \includegraphics[width=1.1\linewidth]{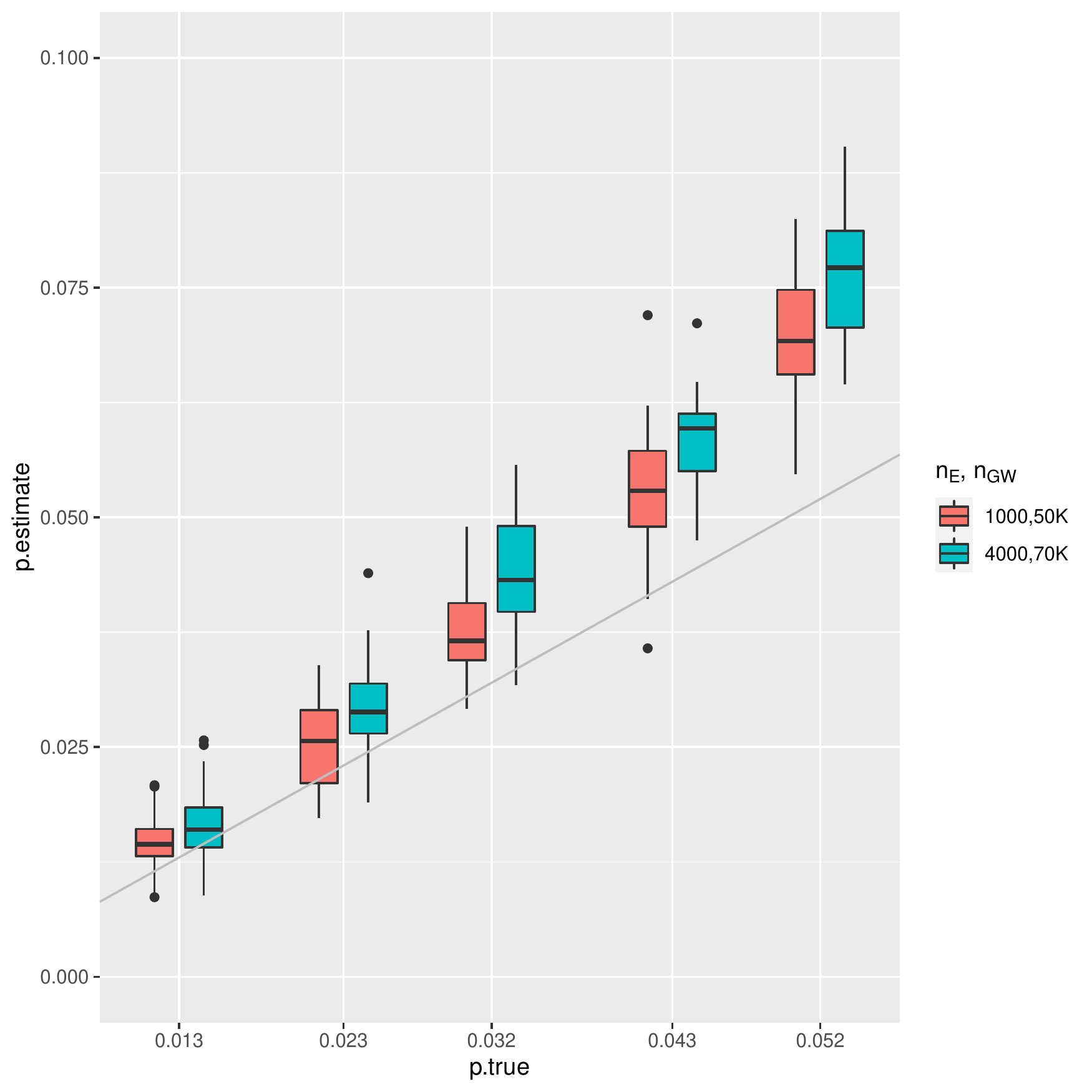}
  \caption{$h^2 = 10\%, 20\%, 30\%, 40\%, 50\%$}
  \label{fig:sub1}
\end{subfigure} \\
\begin{subfigure}{0.45\textwidth}
  \centering
    \includegraphics[width=1\linewidth]{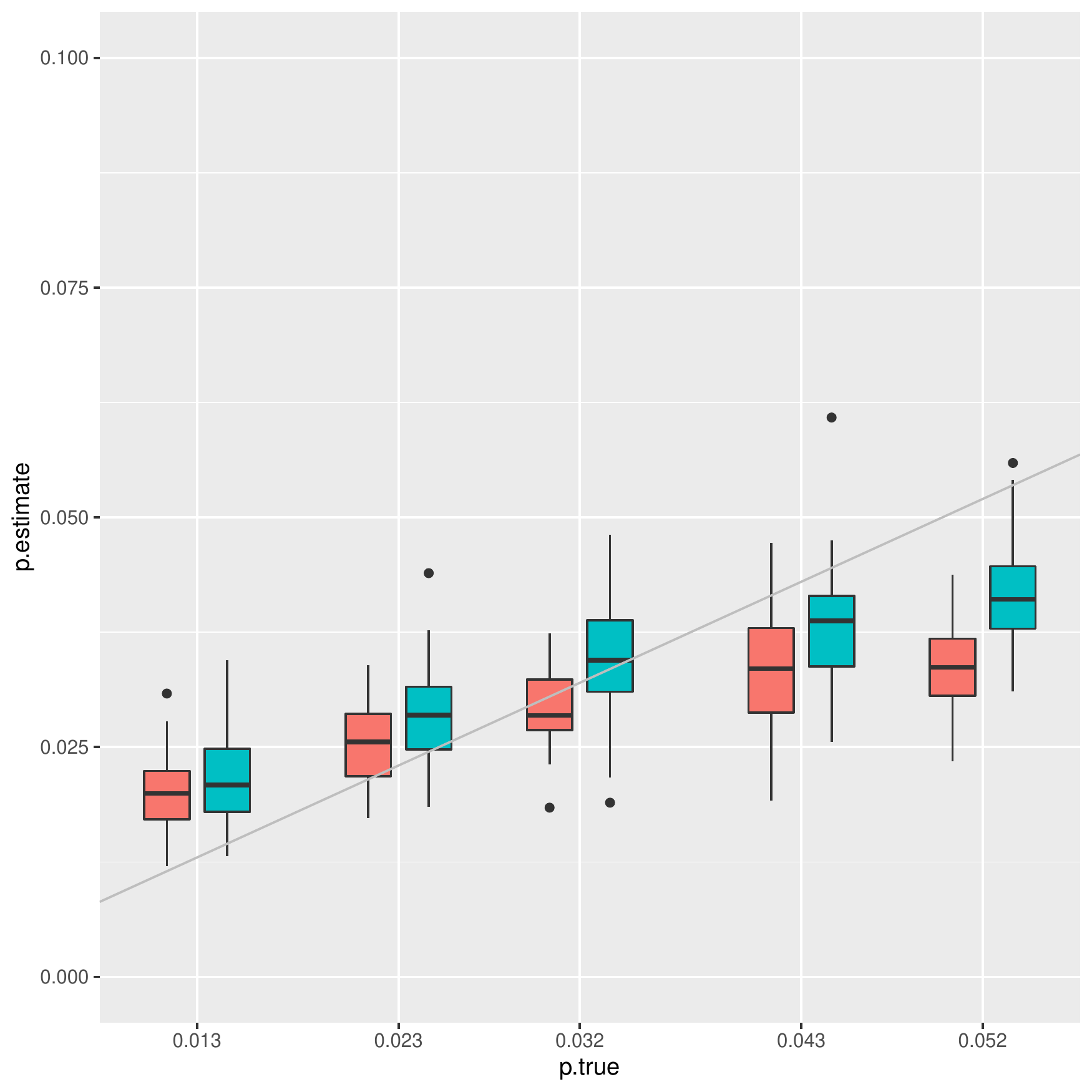}
  \caption{$h^2 = 20\%$}
  \label{fig:sub1}
\end{subfigure}%
\begin{subfigure}{0.45\textwidth}
  \centering
  \includegraphics[width=1.1\linewidth]{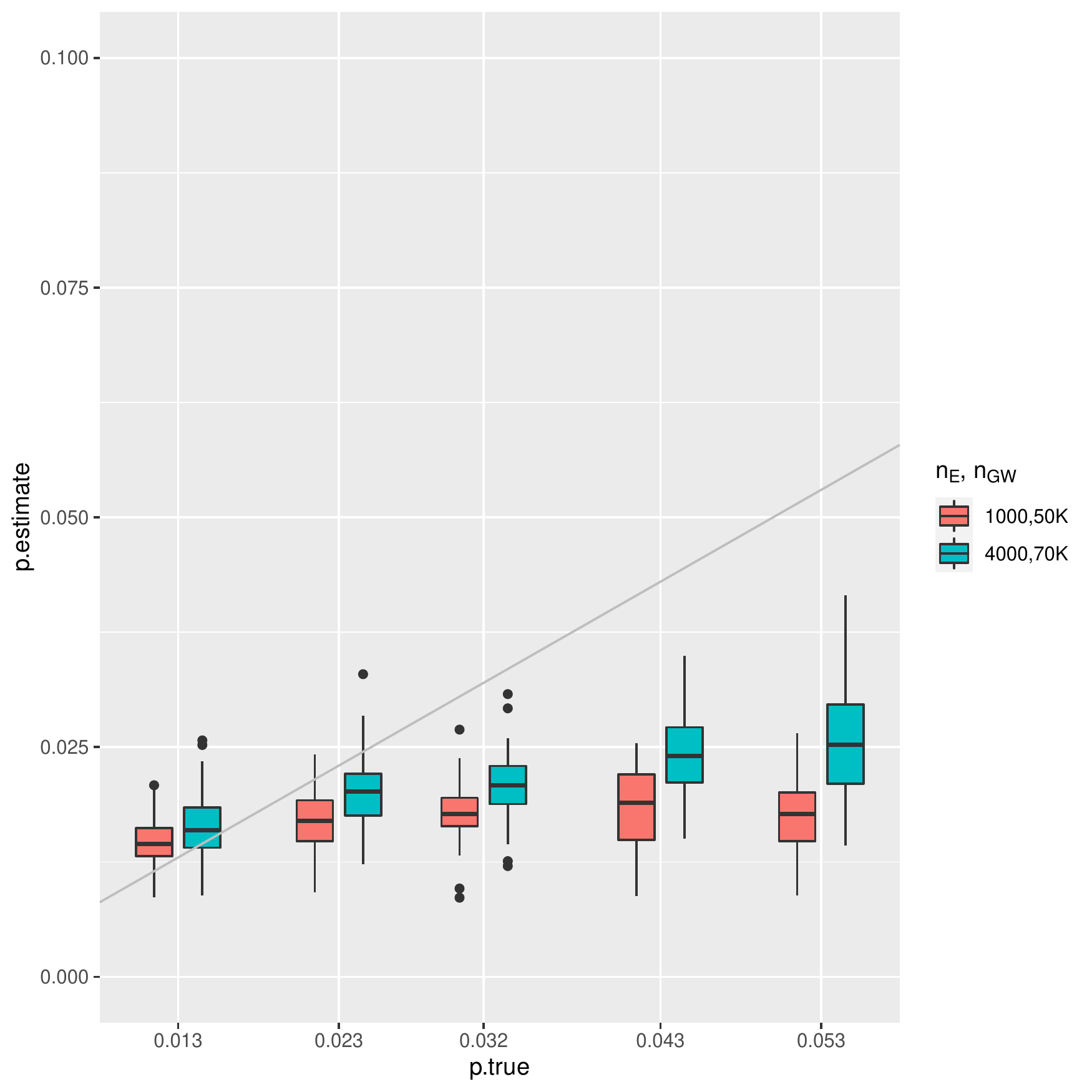}
  \caption{$h^2 = 10\%$}
  \label{fig:sub2}
\end{subfigure}
\caption{Estimated polygenicity when the sample size increases for both the reference panel expression and GWAS data. We increase reference panel sample size from 1000 to 4000, and GWAS sample size from $50K$ to $70K$. The trait heritability due to genetic component of expressions are $10\%,20\%,30\%,40\%,50\%$ for $1\%,2\%,3\%,4\%,5\%$ non-null genes, respectively (a), fixed at $20\%$ (b), fixed at $10\%$ (c).}
\label{fig:est-p-increase-nE-nGW}
\end{figure}

\clearpage

\begin{table}[]
\centering
\caption{Percentage of relative bias produced by the q-value approach while estimating $p$ in various simulation scenarios. The estimated $p$ is based on standard TWAS. $n_E$ and $n_{GW}$ denote the sample sizes of the reference panel expression and GWAS data. We use the following formula of relative bias: $\frac{\text{estimated } p \mbox{ } - \mbox{ } \text{true } p}{\text{true } p} \times 100$}
\label{tb:qvalue}
\begin{tabular}{cccccc}
\hline
        &               & $n_E = 1000$ & $n_E=1000$ & $n_E=4000$ & $n_E=4000$ \\
$p$   & $h^2$   & $n_{GW} = 50K$ & $n_{GW}=70K$ & $n_{GW}=50K$ & $n_{GW}=70K$ \\ \hline
1\% & 10\% & 1149            & 1202            & 979             & 1067            \\
2\% & 20\% & 670             & 699             & 574             & 612             \\
3\% & 30\% & 459             & 493             & 441             & 445             \\
4\% & 40\% & 367             & 400             & 355             & 372             \\
5\% & 50\% & 349             & 340             & 296             & 330             \\
1\% & 20\% & 1219            & 1258            & 971             & 1131            \\
3\% & 20\% & 446             & 480             & 403             & 420             \\
4\% & 20\% & 324             & 348             & 306             & 313             \\
5\% & 20\% & 267             & 277             & 212             & 251             \\
2\% & 10\% & 604             & 654             & 567             & 576             \\
3\% & 10\% & 413             & 438             & 425             & 389             \\
4\% & 10\% & 302             & 307             & 260             & 260             \\
5\% & 10\% & 246             & 241             & 179             & 229             \\ \hline
\end{tabular}
\end{table}

\clearpage

\begin{table}[]
\centering
\caption{Subset of non-null genes for Asthma identified by {\em polygene}.}
\label{tb:subset-asthma}
\begin{tabular}{cc}
\hline
CHR & Gene     \\ \hline
1   & QSOX1    \\
7   & PDK4     \\
10  & TMEM180  \\
17  & CASC3    \\
17  & RAPGEFL1 \\ \hline
\end{tabular}
\end{table}

\begin{table}[]
\centering
\caption{Subset of non-null genes for LDL identified by {\em polygene}.}
\label{tb:subset-LDL}
\begin{tabular}{cc}
\hline
CHR & Gene         \\ \hline
1   & CELSR2       \\
2   & AC009404.2   \\
3   & GNL3         \\
3   & RP11-299J3.8 \\
11  & SOX6         \\
19  & CTB-50L17.9  \\ \hline
\end{tabular}
\end{table}

\begin{table}[]
\centering
\caption{Subset of non-null genes for HDL identified by {\em polygene}.}
\label{tb:subset-HDL}
\begin{tabular}{cccc}
\hline
CHR & Gene                   & CHR & Gene          \\ \hline
1   & CELSR2                 & 10  & AGAP4         \\
1   & CTSS                   & 11  & SOX6          \\
1   & CERS2                  & 11  & HSD17B12      \\
1   & RP11-316M1.12          & 11  & RP11-613D13.5 \\
3   & PPM1M                  & 12  & GLIPR1L2      \\
3   & GNL3                   & 12  & RP11-585P4.5  \\
3   & RP5-966M1.6            & 14  & GSTZ1         \\
3   & RP11-234A1.1           & 15  & RPL9P25       \\
4   & RP11-33B1.4            & 15  & WHAMM         \\
5   & ARHGEF28               & 17  & NR1D1         \\
5   & ERAP2                  & 17  & WIPI1         \\
7   & AC091729.9             & 19  & ZNF846        \\
7   & PILRB                  & 19  & SSBP4         \\
7   & STAG3L5P-PVRIG2P-PILRB & 19  & ITPKC         \\
7   & PILRA                  & 19  & ETHE1         \\
7   & RP11-134L10.1          & 19  & ZNF211        \\
8   & SNX16                  & 22  & BCR           \\ \hline
\end{tabular}
\end{table}

\clearpage

\begin{table}[]
\centering
\caption{Subset of non-null genes for triglycerides identified by {\em polygene}.}
\label{tb:subset-triglycerides}
\begin{tabular}{cccc}
\hline
CHR & Gene          & CHR & Gene          \\ \hline
1   & CELSR2        & 7   & RP11-274B21.3 \\
1   & CTSS          & 9   & PMPCA         \\
1   & CERS2         & 10  & APBB1IP       \\
1   & RP11-316M1.12 & 10  & ALDH18A1      \\
3   & PPM1M         & 11  & SPTY2D1-AS1   \\
3   & GNL3          & 11  & RP11-613D13.5 \\
4   & RP11-33B1.4   & 11  & CEP57         \\
4   & RP11-33B1.1   & 14  & GSTZ1         \\
7   & RP11-274B21.1 & 15  & RPL9P25       \\
7   & RP11-274B21.2 & 15  & WHAMM         \\
7   & RP11-274B21.4 & 16  & PDXDC2P       \\
7   & AC018638.1    & 19  & CTB-50L17.9   \\ \hline
\end{tabular}
\end{table}

\clearpage

\begin{table}[]
\centering
\caption{Subset of non-null genes for BMI identified by {\em polygene}.}
\label{tb:subset-BMI}
\begin{tabular}{cccccccc}
\hline
CHR & Gene         & CHR & Gene          & CHR & Gene          & CHR & Gene          \\ \hline
1   & CROCC        & 3   & PPP2R3A       & 10  & RP11-179B2.2  & 16  & NLRC3         \\
1   & MST1L        & 3   & RP11-731C17.2 & 10  & ARL3          & 16  & SLX4          \\
1   & RP11-108M9.4 & 3   & RP11-85F14.5  & 10  & LHPP          & 16  & CTD-2033A16.3 \\
1   & RP11-108M9.5 & 3   & COMMD2        & 11  & TMEM9B-AS1    & 16  & NOB1          \\
1   & PPIE         & 3   & KLHL24        & 11  & ARL14EP       & 16  & PDPR          \\
1   & ACADM        & 3   & YEATS2-AS1    & 11  & HSD17B12      & 16  & RP11-296I10.3 \\
1   & ECM1         & 3   & ABCC5         & 11  & MRPL11        & 17  & PNPO          \\
1   & IRF6         & 4   & GRK4          & 11  & CTD-3074O7.12 & 17  & RP11-6N17.9   \\
1   & RP3-434O14.8 & 4   & CISD2         & 11  & MRPL21        & 17  & CDK5RAP3      \\
2   & AC092159.2   & 4   & RP11-33B1.4   & 11  & IGHMBP2       & 17  & CALCOCO2      \\
2   & FAM228B      & 4   & RP11-33B1.1   & 11  & ARHGAP20      & 17  & NARF          \\
2   & TRMT61B      & 4   & ELF2          & 11  & TTC12         & 19  & EVI5L         \\
2   & GGCX         & 5   & RNF180        & 11  & HMBS          & 19  & LRRC25        \\
2   & MRPS9        & 7   & PMS2P5        & 12  & PPHLN1        & 19  & UPF1          \\
2   & PHOSPHO2     & 7   & LAMTOR4       & 12  & RP11-328C8.4  & 19  & ARMC6         \\
2   & KLHL23       & 7   & AP1S1         & 12  & PRICKLE1      & 19  & PAF1          \\
2   & TTC30A       & 7   & EXOC4         & 12  & RP11-394J1.2  & 20  & GSS           \\
3   & EAF1         & 8   & FAM86B3P      & 12  & ZNF605        & 20  & MYH7B         \\
3   & CCDC71       & 8   & NTAN1P2       & 13  & MIPEP         & 20  & YWHAB         \\
3   & GPX1         & 8   & DPY19L4       & 14  & FAM177A1      & 21  & PSMG1         \\
3   & FAM212A      & 9   & C9orf40       & 14  & ZFYVE1        & 21  & BX322557.10   \\
3   & GLYCTK       & 9   & C9orf41       & 15  & TIPIN         & 21  & LINC00205     \\
3   & ITIH4-AS1    & 9   & NMRK1         & 15  & MAP2K5        & 22  & ZDHHC8        \\
3   & IQCB1        & 9   & PRPS1P2       & 16  & CLUAP1        &     &               \\ \hline
\end{tabular}
\end{table}

\begin{table}[]
\centering
\caption{Subset of non-null genes for WHR identified by {\em polygene}.}
\label{tb:subset-WHR}
\begin{tabular}{cccccccc}
\hline
CHR & Gene          & CHR & Gene         & CHR & Gene          & CHR & Gene          \\ \hline
1   & THAP3         & 3   & RP11-85F14.5 & 10  & NOC3L         & 15  & WHAMM         \\
1   & MUL1          & 3   & TIPARP-AS1   & 10  & CACUL1        & 15  & UBE2Q2P1      \\
1   & ZNHIT6        & 3   & LINC00886    & 11  & RP11-613D13.5 & 16  & MIR940        \\
1   & PIGC          & 4   & GRK4         & 11  & RAB30         & 16  & ZNF263        \\
2   & UBXN2A        & 4   & TMEM165      & 11  & TMEM126A      & 16  & RP11-266L9.4  \\
2   & ADCY3         & 4   & AADAT        & 11  & PGR           & 17  & NAGLU         \\
2   & RP11-443B20.1 & 5   & LYSMD3       & 11  & TTC12         & 17  & RP11-400F19.8 \\
2   & HOXD-AS1      & 5   & SIL1         & 11  & VPS11         & 17  & CRHR1-IT1     \\
2   & NDUFS1        & 5   & PCDHB2       & 11  & HMBS          & 17  & CRHR1         \\
2   & AC007383.4    & 6   & RP11-250B2.3 & 12  & POC1B         & 17  & RP11-6N17.6   \\
2   & METTL21A      & 6   & GPR126       & 12  & C12orf52      & 17  & RP11-6N17.10  \\
2   & 2-Sep         & 6   & RP11-545I5.3 & 12  & ZNF605        & 17  & HOXB2         \\
3   & RP11-380O24.1 & 6   & VIP          & 13  & RNASEH2B      & 17  & HOXB-AS1      \\
3   & SETD5-AS1     & 7   & INTS1        & 13  & RNF219        & 17  & HOXB3         \\
3   & TCTA          & 7   & AC092171.4   & 14  & AL132989.1    & 17  & RP11-147L13.8 \\
3   & NICN1         & 7   & SNX10        & 14  & PCNX          & 18  & KATNAL2       \\
3   & NICN1-AS1     & 7   & AC004540.4   & 14  & MARK3         & 19  & ANKRD24       \\
3   & RNF123        & 7   & BCL7B        & 14  & APOPT1        & 19  & XAB2          \\
3   & GNL3          & 7   & MLXIPL       & 14  & RP11-73M18.8  & 19  & ZNF100        \\
3   & NEK4          & 7   & AP1S1        & 14  & XRCC3         & 19  & ZNF507        \\
3   & TMEM110       & 7   & RP11-514P8.8 & 14  & PPP1R13B      & 19  & PINLYP        \\
3   & WDR52         & 7   & POLR2J3      & 15  & GABPB1-AS1    & 20  & TRPC4AP       \\
3   & FAM86JP       & 7   & CPED1        & 15  & CTD-3110H11.1 & 20  & EDEM2         \\
3   & SLC41A3       & 7   & EXOC4        & 15  & DIS3L         & 20  & PROCR         \\
3   & RP11-124N2.1  & 9   & DNLZ         & 15  & TIPIN         & 20  & MMP24         \\
3   & PCCB          & 10  & ZCCHC24      & 15  & MAP2K5        & 20  & UQCC1         \\
    &               &     &              &     &               & 22  & THOC5         \\ \hline
\end{tabular}
\end{table}

\begin{table}[]
\centering
\caption{Subset of non-null genes for height identified by {\em polygene}.}
\label{tb:subset-height1}
\begin{tabular}{cccccccc}
\hline
CHR & Gene           & CHR & Gene          & CHR & Gene          & CHR & Gene         \\ \hline
1   & PGD            & 1   & EGLN1         & 3   & NICN1         & 4   & UBE2D3       \\
1   & NBPF1          & 2   & SH3YL1        & 3   & NICN1-AS1     & 4   & RP11-33B1.1  \\
1   & CROCC          & 2   & ADI1          & 3   & FAM212A       & 5   & C5orf22      \\
1   & MST1L          & 2   & AC142528.1    & 3   & NEK4          & 5   & NSA2         \\
1   & RP11-108M9.4   & 2   & UBXN2A        & 3   & ITIH4-AS1     & 5   & DHFR         \\
1   & RP11-108M9.5   & 2   & PFN4          & 3   & RP5-966M1.6   & 5   & MSH3         \\
1   & PADI2          & 2   & HMGB1P31      & 3   & TMEM110       & 5   & GIN1         \\
1   & DDOST          & 2   & RP11-493E12.2 & 3   & RFT1          & 5   & PRR16        \\
1   & EIF4G3         & 2   & GGCX          & 3   & LNP1          & 5   & ALDH7A1      \\
1   & PIGV           & 2   & ZNF514        & 3   & PCCB          & 5   & SRA1         \\
1   & FNDC5          & 2   & C2orf49       & 3   & CEP70         & 5   & PCDHB16      \\
1   & MEAF6          & 2   & C2orf40       & 3   & ATP1B3        & 5   & PANK3        \\
1   & RP11-767N6.7   & 2   & ERCC3         & 3   & PCOLCE2       & 6   & RNF144B      \\
1   & IPP            & 2   & AC093388.3    & 3   & PLOD2         & 6   & PRICKLE4     \\
1   & MAST2          & 2   & PPIL3         & 3   & RSRC1         & 6   & MED20        \\
1   & GBP3           & 2   & NOP58         & 3   & RP11-538P18.2 & 6   & CD2AP        \\
1   & CELSR2         & 2   & KCNE4         & 3   & YEATS2-AS1    & 6   & HMGN3        \\
1   & CHD1L          & 2   & TRAF3IP1      & 4   & Z95704.2      & 6   & RP11-250B2.6 \\
1   & RP11-337C18.10 & 2   & NDUFA10       & 4   & HTT           & 6   & DOPEY1       \\
1   & ACP6           & 2   & HDLBP         & 4   & QDPR          & 6   & SMIM8        \\
1   & POGZ           & 2   & 2-Sep         & 4   & DANCR         & 6   & C6orf163     \\
1   & FMO4           & 3   & XPC           & 4   & SRD5A3        & 6   & FRK          \\
1   & NCF2           & 3   & LSM3          & 4   & SRD5A3-AS1    & 6   & NUS1         \\
1   & CDC42BPA       & 3   & CRTAP         & 4   & REST          & 6   & TULP4        \\
1   & SNAP47         & 3   & AMT           & 4   & UBA6          & 6   & IGF2R        \\ \hline
\end{tabular}
\end{table}

\begin{table}[]
\centering
\caption{Subset of non-null genes for height identified by {\em polygene}.}
\label{tb:subset-height2}
\begin{tabular}{cccccccc}
\hline
CHR & Gene                   & CHR & Gene          & CHR & Gene          & CHR & Gene          \\ \hline
7   & EIF3B                  & 8   & CSGALNACT1    & 10  & PAPSS2        & 12  & C12orf23      \\
7   & DPY19L1P1              & 8   & BRF2          & 10  & NOC3L         & 12  & MMAB          \\
7   & KBTBD2                 & 8   & ADAM9         & 10  & CWF19L1       & 12  & DDX55         \\
7   & GUSB                   & 8   & RPS20         & 10  & TMEM180       & 12  & EIF2B1        \\
7   & GS1-124K5.11           & 8   & DSCC1         & 10  & PRDX3         & 12  & TCTN2         \\
7   & GS1-124K5.12           & 8   & WDYHV1        & 10  & PLEKHA1       & 13  & C1QTNF9B-AS1  \\
7   & KCTD7                  & 9   & RP11-112J3.16 & 10  & JAKMIP3       & 13  & COG6          \\
7   & RP11-458F8.2           & 9   & TMEM8B        & 11  & LDHA          & 14  & CTD-2002H8.2  \\
7   & TYW1                   & 9   & HRCT1         & 11  & TSG101        & 14  & FLVCR2        \\
7   & RP11-166O4.5           & 9   & RP11-522I20.3 & 11  & MPPED2        & 14  & RP11-747H7.3  \\
7   & POM121B                & 9   & GKAP1         & 11  & AP000442.1    & 14  & TRIP11        \\
7   & CYP51A1                & 9   & RMI1          & 11  & MS4A14        & 14  & RP11-529H20.6 \\
7   & PEX1                   & 9   & CDC20P1       & 11  & CTSF          & 14  & ATXN3         \\
7   & ZSCAN21                & 9   & FKBP15        & 11  & TPCN2         & 14  & NDUFB1        \\
7   & PILRB                  & 9   & CDC26         & 11  & SYTL2         & 14  & ZFYVE21       \\
7   & STAG3L5P-PVRIG2P & 9   & GOLGA1        & 11  & MED17         & 15  & C15orf41      \\
   & -PILRB &    &         &   &          &   &       \\
7   & PILRA                  & 9   & PPP6C         & 11  & AP001877.1    & 15  & DIS3L         \\
7   & TSC22D4                & 9   & GAPVD1        & 11  & FDX1          & 15  & RP11-352G18.2 \\
7   & GIGYF1                 & 9   & DNLZ          & 12  & WBP11         & 15  & TIPIN         \\
7   & TRIP6                  & 9   & CARD9         & 12  & RP11-967K21.1 & 15  & RPL9P25       \\
7   & TSPAN33                & 10  & AKR1C2        & 12  & RP11-996F15.2 & 15  & SNAPC5        \\
7   & RP11-286H14.6          & 10  & PHYH          & 12  & RP11-611E13.2 & 15  & SMAD3         \\
7   & RP4-800G7.2            & 10  & KIF5B         & 12  & MRPL42        & 15  & AP3B2         \\
7   & NOM1                   & 10  & DNA2          & 12  & CCDC53        & 15  & RP11-182J1.14 \\
8   & MFHAS1                 & 10  & RPS24         & 12  & CKAP4         & 15  & RN7SL417P     \\ \hline
\end{tabular}
\end{table}

\begin{table}[]
\centering
\caption{Subset of non-null genes for height identified by {\em polygene}.}
\label{tb:subset-height3}
\begin{tabular}{cccccccc}
\hline
CHR & Gene         & CHR & Gene         & CHR & Gene       & CHR & Gene         \\ \hline
15  & WDR73        & 17  & EFCAB13      & 19  & SUGP1      & 20  & GGT7         \\
15  & ALPK3        & 17  & MRPL45P2     & 19  & MAU2       & 20  & ACSS2        \\
15  & DET1         & 17  & RP11-6N17.4  & 19  & ZNF714     & 20  & GSS          \\
15  & POLG         & 17  & AC003665.1   & 19  & ZNF738     & 20  & MMP24-AS1    \\
15  & PEX11A       & 17  & PNPO         & 19  & ZNF493     & 20  & MMP24        \\
16  & BRICD5       & 17  & RP11-6N17.6  & 19  & ZNF429     & 20  & RP4-614O4.11 \\
16  & ECI1         & 17  & RP5-890E16.2 & 19  & ZNF100     & 20  & EIF6         \\
16  & MIR940       & 17  & HOXB3        & 19  & ZNF43      & 20  & UQCC1        \\
16  & NTAN1        & 17  & COX11        & 19  & ZNF507     & 20  & YWHAB        \\
16  & NPIPA5       & 17  & BZRAP1       & 19  & GPATCH1    & 20  & SNX21        \\
16  & ABCC6        & 17  & LRRC37A16P   & 19  & COX6B1     & 20  & PREX1        \\
16  & TANGO6       & 17  & CCDC40       & 19  & WDR62      & 20  & B4GALT5      \\
16  & PHLPP2       & 17  & GAA          & 19  & TBCB       & 21  & PIGP         \\
16  & RFWD3        & 17  & RAB40B       & 19  & LINC00665  & 21  & TTC3         \\
17  & ZNF286B      & 17  & FN3KRP       & 19  & ITPKC      & 21  & UBE2G2       \\
17  & RP11-173M1.8 & 18  & POLI         & 19  & PHLDB3     & 21  & POFUT2       \\
17  & TP53I13      & 18  & CNDP2        & 19  & ETHE1      & 22  & UFD1L        \\
17  & RAPGEFL1     & 19  & MPND         & 19  & ZNF575     & 22  & AC000068.10  \\
17  & HSD17B1P1    & 19  & AC007292.3   & 19  & SMG9       & 22  & AP000350.4   \\
17  & DCAKD        & 19  & CHAF1A       & 19  & DMWD       & 22  & DDT          \\
17  & PLCD3        & 19  & CTB-50L17.5  & 19  & FUZ        & 22  & CTA-445C9.15 \\
17  & PLEKHM1      & 19  & CTB-50L17.9  & 19  & ZNF667-AS1 & 22  & MTMR3        \\
17  & LRRC37A4P    & 19  & KRI1         & 20  & C20orf194  & 22  & PLA2G6       \\
17  & CRHR1-IT1    & 19  & SLC44A2      & 20  & CBFA2T2    & 22  & DMC1         \\
17  & CRHR1        & 19  & SSBP4        & 20  & MAP1LC3A   & 22  & TOMM22       \\ \hline
\end{tabular}
\end{table}

\end{document}